\documentclass[aps,10pt,prb,twocolumn,superscriptaddress,showpacs,floats,floatfix,preprintnumbers,amsmath,amssymb]{revtex4-1}
\usepackage[T1]{fontenc}
\usepackage[latin9]{inputenc}
\usepackage[USenglish,american]{babel}
\usepackage{amsmath}
\usepackage{amssymb}
\usepackage{wasysym}
\usepackage{graphicx}
\usepackage{tabularx}
\usepackage{xcolor}
\usepackage{bm}          
\usepackage{pifont} 
\usepackage{dcolumn}
\usepackage{float}
\usepackage{natbib}
\usepackage{enumerate}

\usepackage{subfigure,textcomp,epsfig,siunitx}
\usepackage{array}
\usepackage{longtable}
\usepackage{bm}
\usepackage{booktabs}

\usepackage[colorinlistoftodos]{todonotes}

\newcommand{\GrazPhys}{Institute of Theoretical and Computational Physics,
Graz University of Technology, NAWI Graz, 8010 Graz, Austria}
\newcommand{\UniRoma}{Dipartimento di Fisica, Universit\`a di Roma La Sapienza, Piazzale Aldo Moro 5, I-00185 Roma, Italy}
\newcommand{\KnowCenter}{Know-Center GmbH, Research Center for Data-Driven Business \& Big Data Analytics, Inffeldgasse 13, 8010 Graz, Austria}

\begin{document}


\title{Importance of feature engineering and database selection in a machine learning model: A case
study on carbon crystal structures}


\author{Franz M. Rohrhofer} \affiliation{\GrazPhys} \affiliation{\KnowCenter}
\author{Santanu Saha} \email{santanu.saha@tugraz.at} \affiliation{\GrazPhys}
\author{Simone Di Cataldo} \affiliation{\GrazPhys} \affiliation{\UniRoma}
\author{Bernhard C. Geiger} \affiliation{\KnowCenter}
\author{Wolfgang von der Linden} \affiliation{\GrazPhys}
\author{Lilia Boeri} \affiliation{\UniRoma}

\date{\today}

\begin{abstract}
Drive towards improved performance of machine learning models has led to the
creation of complex features representing a database  of condensed matter systems.
The complex features, however, do not offer an intuitive explanation on which
physical attributes do improve the performance. The effect of the database  on
the performance of the trained model is often neglected.
In this work we seek to understand in depth 
the effect that the choice of features and the properties of the database have on a machine 
learning application. In our experiments, we consider the complex phase space of carbon as a test case,
for which we use a set of simple, human understandable and cheaply computable features 
for the aim of predicting the total energy of the crystal structure. 
 Our study shows that (i) the performance of the machine learning model varies
depending on the set of features and the database, (ii) is not transferable to
every structure in the phase space and (iii) depends on how well structures are
represented in the database.
\end{abstract}

\maketitle

\section{Introduction}\label{sec:INTRODUCTION}
The use of machine learning (ML) algorithms in the study of condensed matter systems
has emerged as a new paradigm for study and discovery of new functional materials
~\cite{schleder2019dft}. ML makes it possible to uncover the trends in huge datasets
and to make accurate predictions of material properties at a cheaper cost than calculations 
from first principles. The field has already witnessed real-life applications in the 
field of drug discovery~\cite{lo2018machine}, catalysis~\cite{kitchin2018machine}, and
solar-based technologies~\cite{sahu2018toward}, to name a few.

In order to build a ML model that predicts a target property of a condensed matter
systems, one requires: (i) a set of features to represent the structures, (ii) a
database of structures for training, and (iii) a suitable ML algorithm.
The performance of the trained ML model is measured by the quality of its prediction
through different tests and error estimates. 
The key interest of the ML studies in this field had
been the improvement of performance through engineered features/algorithms or their
application to different systems of interest. However, there exists almost no
literature which sheds light on how the individual constituents of a ML model
influence the outcome~\cite{haghighatlari2020learning}.

An in-depth study of the inter-dependency of different factors on the performance of a ML
model is not only crucial for better understanding, but would also provide a clear, practical
guideline for constructing better performing ML models. We investigate this aspect by carrying 
out systematic tests on the features and database through performance of the trained ML model.
We considered the Kernel Ridge Regression (KRR) as the choice of ML model.

We consider the total energy of carbon crystal structures as target property that 
our ML model shall predict. The database of carbon crystal structures used have been
generated using crystal structure prediction methods at the level of density functional theory.


The first crucial component required to build a ML model for condensed matter systems is a set of
features representing the database. Ideally, the set of features used for representing any condensed matter
system should be: (i) unique, (ii) invariant under rotation, translation and permutation of atoms
and (iii) continuous. Generally, features capturing relevant information of a system associated
to a target property often lead to improved performance. However, it is difficult to construct
features that capture this relevant information and simultaneously satisfy the desirable properties
(i)-(iii). As for example, the traditionally used atomic positions and lattice vectors to represent
condensed matter systems in ab-initio calculations fail to satisfy these necessary criteria.

The effort towards the identification of suitable features for condensed matter systems has led
to the development of complex mathematical functions which primarily focus on the local atomic
environment~\cite{behler2011atom,smith2017ani,faber2018alchemical,christensen2020fchl,bartok2013representing,zhu2016fingerprint}. 
A comparative study of the performance given by different features can be found in 
Ref.~\cite{parsaeifard2020assessment}. These features have been tested and applied mostly to
the study of aperiodic systems~\cite{rupp2012fast,hansen2013assessment,montavon2013machine}. 
Direct application of these features to periodic systems may push the requirement of large
database for training and, hence, make it computationally expensive.
Nevertheless, they have been successfully used to study periodic systems
~\cite{rowe2020accurate,deringer2018data,bartok2013representing,engel2018mapping,cheng2020evidence}.

A major drawback of complex features based on the local atomic environment is that they 
do not offer an intuitive understanding. Though this aspect doesn't effect the performance 
of the ML model, this is crucial in understanding the type of structures present in database. 
This understanding will aid in creation of subsets of the database and thereby help 
understanding the influence of database in the performance of ML model.
In this regard, the radial distribution function (RDF), a mathematically simplistic and
semi-intuitive feature, has been a popular choice in application of ML for structure prediction
studies. This advantage in the use of RDF is that it does a decent job in terms of performance with
reduced complexity~\cite{ward2017including,honrao2019machine}. However, the RDF is prone to
loss of information due to binning and averaging. On top, using RDF constructed for every atom 
of the unit cell would directly lead to exponential increase in number of features and hence, the database
required for training. Hence, RDF is not sufficient to attain the
objective of our ML study. Thus, we also include angular distribution function
and a set of scalar valued physically meaningful features.

The influence of the other crucial component of any ML study, the database, is generally not well
studied. In most ML studies, the database is either acquired from different public repositories
~\cite{bergerhoff1983inorganic,belsky2002new,jain2013commentary,saal2013materials}
or generated through the use of different ab-initio based structure prediction algorithms
~\cite{glass2006uspex,lyakhov2013new,pickard2011ab,goedecker2004minima,amsler2010crystal}.
Secondly, irrespective of the source of database, the quality of the calculated quantities still
remains an issue. Database available in open repositories such as Materials Project~\cite{jain2013commentary},
OQMD~\cite{saal2013materials} at present state may or may not have same level of accuracy/precision,
whereas in case of generated structures, the loose settings used for quick evaluation of target
properties not necessarily represent the true performance as this practice induce errors bigger
than the accuracy of the model itself.

On top of all these inconsistencies in the database, it has become regular practice to add/remove
certain sets of structures from the database to improve performance without any proper justification.
Thus, while the database was occassionally adapted to improve performance of the ML model, its influence
in ML model building has not been studied before in the field of condensed matter systems. The main
drive of the ML studies had been to engineer the features to obtain better performance.

For our investigation, we sought to create a database of a system which is complex, providing a large
pool of distinct structures. Among different known complex systems, carbon is perhaps the only
one whose complex chemistry is well understood despite it being polymorphic with a large number of
allotropes. The fact that the complex chemistry of carbon boils down to $sp^1-sp^2-sp^3$ bonds
greatly elucidates the understanding of different carbon allotropes. This dual characteristic of
carbon makes it an ideal test bed for our ML study. 
In order to generate a large pool of diverse structures quickly and efficiently, we employed
crystal structure prediction methods.

Finally, we discuss our choice of ML model.
Different ML models used for the study of condensed matter systems come with their own
set of (dis)advantages. Often, this choice can be guided by the system of interest, the
kind of features used and the size of the database. One of the most popular ML models used for
condensed matter systems is KRR. KRR is a powerful model for nonlinear regression, easy
to optimize and its requirement of data for training is limited. This makes KRR optimal
for its use in condensed matter systems and for our study in particular. Although we have
tested different models such as Ridge Regression, Lasso Regression and Support Vector
Regression, for the course of this work we stick with KRR. We did not take neural
networks~\cite{behler2007generalized}, another popular ML model used in condensed matter
systems, into consideration due to its requirement of a comparatively large database for training.

The work carried out in the manuscript is arranged as follows. In Sec.~\ref{sec:DATA-FEATURES}
the features used for the data representation are introduced. Sec.~\ref{sec:CarbonDB}
discusses the generation and creation of different database of carbon. Sec.~\ref{sec:MLmodel} presents
the ML model used in this work. Sec.~\ref{sec:Results} presents the results of the ML studies
carried out on different combination of database and sets of features. In the end the conclusions of
the studies are presented in Sec.~\ref{sec:conclusion}. The methodology used and details of the
calculations are provided in Appendix~\ref{sec:AppMethodology} and plots of distributions for
different features in Appendix~\ref{sec:AppHistogram}.

\section{Data Representation}\label{sec:DATA-FEATURES}

\begin{figure*}
\centering
\includegraphics[width=1\textwidth]{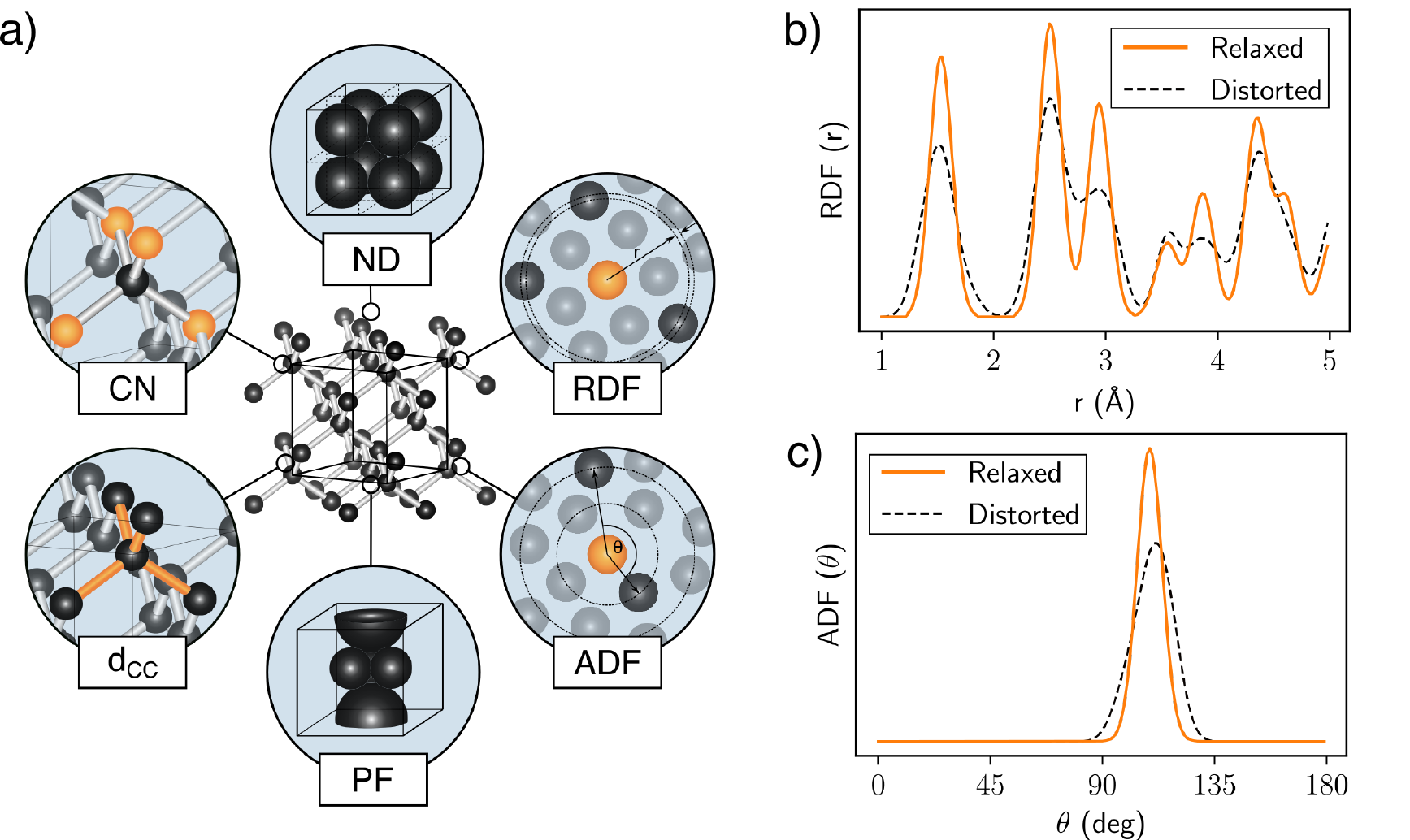}
\caption {\textbf{(a)} The features used for representing a carbon crystal structure in the ML model
are: the number density (ND), coordination number (CN), bond length ($d_{CC}$ in~\AA),
packing fraction (PF), angular distribution function (ADF) and radial distribution function (RDF).
\textbf{(b)} The RDF of relaxed (solid orange) and distorted (dashed black) diamond crystal structure 
using bin size $\Delta r$ = 0.1~\AA, cutoff radius $r_\text{c}$ = 5~\AA~and Gaussian smearing 
$\sigma_\text{RDF}$ = 0.1~\AA. \textbf{(c)} The ADF of relaxed (solid orange) and distorted 
(dashed black) diamond crystal structure using bin size $\Delta\theta$ = 15$^\circ$, 
$\theta$ = 0$^\circ$-180$^\circ$ and Gaussian smearing $\sigma_\text{ADF}$ = 15$^\circ$.}
\label{fig:feature}
\end{figure*}

A schematic representation of the set of features used for our ML study is shown in
Fig.~\ref{fig:feature}. The features shown are for an 8-atom unit cell of diamond.
The features used are: (i) Radial distribution function (RDF), (ii) Angular distribution
function (ADF) and (iii) single geometric descriptors (SGD),
a combination of scalar physically meaningful features. 
These three features are discussed below:

\begin{itemize}
\item \textbf{Radial distribution function (RDF):} \\
The RDF describes the chemical bonding environment of atoms and is a measure of the averaged
local density at distance $r$ compared to the bulk number density. In this work, we adapted
the RDF from \textit{Sch\"utt} et al.~\cite{schutt2014represent} and constructed the RDF
according to
\begin{equation}
g(r)=\frac{V_{\text{uc}}}{N^2}\frac{1}{V_r}\sum_{i=1}^N\sum_j\Theta(d_{ij}-r)\Theta(r+dr-d_{ij})
\label{eq:RDF}
\end{equation}
where $V_{\text{uc}}$ is the volume of the unit cell, $N$ is the number of atoms in the unit cell,
$V_r$ is a spherical shell volume with radius $r$ and infinitesimal shell thickness $dr$, $i$
runs over all the atoms in the unit cell, $j$ runs over all atoms in the extended unit cell,
$d_{ij}$ is the inter-atomic distance between the atoms $i$ and $j$ and $\Theta(d_{ij}-r)$ is
the delta function. For using the RDF as finite-size feature vector, eq.~\eqref{eq:RDF} is
quantized and smoothed as discussed in Appendix~\ref{sec:AppMethodology}. As an example, the
RDF of a relaxed (solid orange) and distorted (dashed black) diamond structure is shown in
Fig.~\ref{fig:feature}(b).

\item \textbf{Angular Distribution Function (ADF):} \\
The ADF captures information on the bond angle distribution among the nearest neighbours.
It is a normalized distribution function of the bond angles made by the atoms 
and is constructed using the relation
\begin{equation}
\begin{split}
\text{ADF}(\theta)= &\frac{1}{A}\sum^N_{i=1}\sum_{<j,k>_{nn}}w(\theta_{jik})\Theta(\theta_{jik}-\theta) \\
&\times\Theta(\theta+d \theta -\theta_{jik})
\end{split}
\label{eq:ADF}
\end{equation}
where $\theta_{jik}$ is the bond angle formed by the atoms $j,i,k$ centered at atom $i$,
$w(\theta_{jik})$ is the weight factor on the bond angle, $\Theta(\theta_{jik}-\theta)$
is the delta function and A is the normalization constant.

The bond angle $\theta_{jik}$ centered at atom $i$ is calculated using 
\begin{equation}
    \theta_{jik} = \frac{180^\circ}{\pi}\cos^{-1}\left( \frac{\vec{r}_{ij}\cdot\vec{r}_{ik}}{\|\vec{r}_{ij}\|\cdot\|\vec{r}_{ik}\|}\right),
\label{eq:theta}
\end{equation}
where $\vec{r}_{ij}=\vec{r}_j-\vec{r}_i$, $\vec{r}_{ik}=\vec{r}_k-\vec{r}_i$ and
$\vec{r}_i$, $\vec{r}_j$, $\vec{r}_k$ are the position vectors of atom $i,j,k$ respectively.

The weight factor $w$($\theta_{jik}$) was introduced to give more importance to nearest
neighbour contributions. It is given by
\begin{equation}
    w(\theta_{jik}) = \frac{1}{\|\vec{r}_{ij}\|^4 \cdot \|\vec{r}_{ik}\|^4} \quad.
\label{eq:wtheta}
\end{equation}
Again, eq.\eqref{eq:ADF} is discretized and smoothed as discussed in Appendix~\ref{sec:AppMethodology}.
As an example, the ADF of a relaxed (solid orange) and distorted (dashed black) diamond structure
is shown in Fig.~\ref{fig:feature}(c).

\item \textbf{Single Geometric Descriptors (SGD):} \\
The RDF is prone to loss of information due to binning and averaging, especially in polymorphic
system such as carbon. Other than this, we also desired a set of features which could provide 
intuitive understanding of the physical/chemical nature of the structures of the database.
This aspect is crucial in understanding the influence of database on the performance of the ML model.
Therefore, we selected a collection of different scalar quantities - average coordination
number (CN), average inter-atomic bond distance ($d_{CC}$) (\AA), number-density (ND) (\AA$^{-3}$)
and packing-fraction (PF). The ND represents the number of atoms per unit volume, whereas PF
represents the ratio of the volume of all the individual atoms to that of the volume of the
unit cell, based on a hard sphere model of the atom. These quantities are bundled under single
geometric descriptors (SGD). These SGD features used along with RDF and ADF are shown in the
schematic diagram in Fig.~\ref{fig:feature}(a).

We considered the mean CN and mean $d_{CC}$ of first nearest neighbour ($CN_{1}$, $d_{1,CC}$)
and second nearest neighbour ($CN_{2}$, $d_{2,CC}$) to be part of SGD. As the carbon structures 
used in the study can have different fraction of $sp^1-sp^2-sp^3$ bonds with large deviation in
bond-distances, only the standard deviations of CN and $d_{CC}$ of the first nearest neighbour
were considered.
\end{itemize}

\begin{table}[!htb]
\caption{List of features, their type, number of features and kind of information they 
capture- local or global. The symbols (\ding{52}) and (\ding{55}) represent whether or not a
feature captures local/global information.}
\begin{tabular}{cccccc} 
\hline 
Feature & Type & Parameters & \# of & Local & Global \\ [0.5ex] 
           &      &            & features &  &   \\
\hline\hline 
$CN_{1}$ & SGD &  & 2 & \ding{52} & \ding{55} \\
(mean, std. dev) & & & & & \\
\hline
$CN_2$ & SGD &  & 1 & \ding{52} & \ding{55} \\
(mean) & & & & & \\
\hline 
$d_\text{1,CC}$ & SGD &  & 2 & \ding{52} &  \ding{55} \\
(mean, std. dev) & & & & & \\
\hline 
$d_\text{2,CC}$ & SGD &  & 1 & \ding{52} & \ding{55} \\
(mean) & & & & & \\
\hline 
ND & SGD &   & 1 & \ding{55} & \ding{52} \\
\hline 
PF &  SGD &  & 1 & \ding{55} & \ding{52} \\ 
\hline 
RDF & RDF & $\Delta r$ = 0.1 \AA & ~40 & \ding{52} & \ding{55} \\ 
    &     &     $r_c$= 1~\AA - 5~\AA &  &  & \\
\hline 
ADF & ADF & $\Delta\theta$= 15$^\circ$ & 12 & \ding{52} & \ding{55}  \\
    &     &     $\theta$= 0$^\circ$-180$^\circ$ &  &  & \\
\hline
\end{tabular}
\label{tab:feature}
\end{table}
The detailed information about the features discussed above, i.e. their type, parameters for
discretization and defining the range, their number of features and kind of information they
represent (local or global) are enlisted in Tab.~\ref{tab:feature}. In total, for a given crystal
structure, one would need just 60 (8 SGD + 40 RDF + 12 ADF) features to represent it in the ML model. 
Details on the construction of features are discussed in the Appendix~\ref{sec:AppMethodology}. 

\section{Database  of Carbon Structures}\label{sec:CarbonDB}

\begin{figure*}[!htb]
\centering
\includegraphics[width=2.0\columnwidth,angle=0]{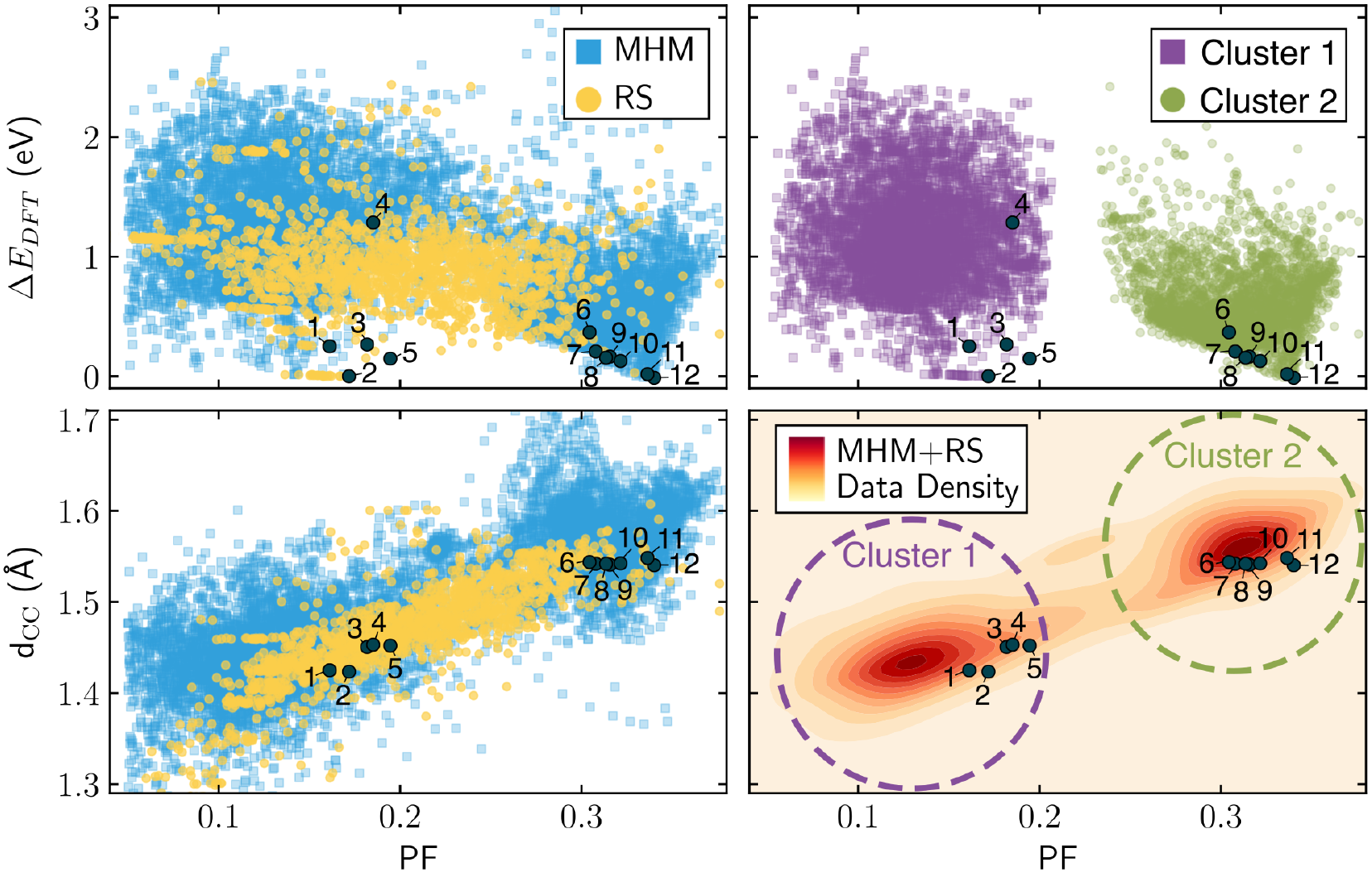}
\caption{Scatter plot of carbon crystal structures as function of packing fraction (PF)
vs.\ first nearest neighbour distance $d_{CC}$ (\AA) in the lower left panel and as a
function of PF vs. energy $\Delta$E$_{DFT}$ (eV/atom) w.r.t.\  graphite in upper left
panel. The blue squares and yellow circles in the scatter plot represent data points
obtained from minima hopping (MHM) and random sampling (RS) methods, respectively.
Same data points grouped into cluster 1 (violet square) and cluster 2 (green
circle) are shown in the right top (PF vs. $\Delta$E$_{DFT}$) and as density plot in
right bottom panel (PF vs. $d_{CC}$). The data points which are outside the circle of
Cluster 1 and Cluster 2 are removed in these plots. The numbers adjacent to the black
circle in the plots represent the position of the following carbon allotropes: (1) 
Haeckelite, (2) Graphite, (3) CHC, (4) K4, (5) ING, (6) D-Carbon, (7) BCT, (8) W-Carbon,
(9) M-Carbon, (10) Z-Carbon, (11) Lonsdalite and (12) Diamond}.
\label{fig:CarbonDB}
\end{figure*}

A diverse training database with a good representation of the phase space of interest is essential
for the development of a robust and universal ML model. In our study, we exploited the 
crystal structure prediction methods to generate a diverse crystal structure database of carbon.
We used a combination of random-structure (RS) and molecular dynamics approaches. The generation
of database in the RS approach was achieved through universal structure predictor: evolutionary 
crystallography (USPEX)~\cite{glass2006uspex,lyakhov2013new} random structure generator, whereas 
the molecular dynamics approach through the minima hopping method
(MHM)~\cite{goedecker2004minima,amsler2010crystal}.

The two methods generate structures with complementary characteristics. In RS
approach, the random structures generated based on certain constraints 
(space-group, bond distances) are geometrically relaxed. These relaxed structures represent
the local minima of the phase space. On contrary, the MHM searches the local minima through
a molecular dynamics trajectory based search starting from structural prototypes to sample
the neighbouring phase space. The structures generated in the molecular dynamics trajectory
represent intermediate structures of the phase space. For our study, we considered both the
local minima and intermediate structures as part of the MHM database. Hence, the structures in
RS approach, i.e. RS database contains only the local minima whereas the local minima and
intermediate structures constitutes the MHM database. In order to retain structural complexity
for our studies without any significant increase in computational costs, we limited the
generation of crystal structures in both methods to 8 atoms per unit cell.

Presence of redundant-unphysical structures in the RS database and MHM database
limits the performance of the ML model. Hence, duplicate and unphysical structures were
removed through the Oganov fingerprint distance~\cite{valle2010crystal} and PF respectively.
Structures with PF $\leqslant$ 0.05 were considered unphysical, i.e.\ loosely packed
structures. After this screening, the final number of structures obtained from both database
was 11500, where 1500 structures were from RS database and 10000 structures from MHM database. 
The relatively large number of structures in MHM database as compared to RS database is due 
to the inclusion of intermediate structures in MHM database. The combination of 11500
structures coming from MHM and RS database is termed as "MHM+RS database" throughout the manuscript.

We selected $10\%$ of the crystal structures from the entire database as a holdout set which was
used for measuring the model accuracy after the model and features were optimized. Since
the holdout set has its origin in the same structure generation as the database used for training
the model, its use as a final test set may not necessarily represent the overall accuracy of
the ML model. To create an independent test set, we additionally selected 12 well-known carbon
allotropes as mentioned in the caption of Fig.~\ref{fig:CarbonDB} and in Appendix
\ref{sec:AppMethodology}.
The choice of the allotropes is motivated by the fact that they have distinct structural
motifs with different combination of $sp^2-sp^3$ bonds and yet can be represented in a 
8-atom unit cell. None of these allotropes were included in the training data.

All the 11500 structures obtained using RS (yellow circles) and MHM (blue square) and
the 12 allotropes (black circles) in the test set are shown as a function of PF vs.
$d_{CC}$ in the lower left and as a function of PF vs. energy per atom in eV w.r.t.\ graphite
$\Delta$E$_{DFT}$ in the upper left panel of Fig~\ref{fig:CarbonDB}. Most of the 1500 RS structures
are concentrated in the range 0.10 $\leqslant$ PF $\leqslant$ 0.30 with few points scattered
above and below the range. In contrast, the data points of MHM structures are spread over the
whole range of PF, 0.05 $\leqslant$ PF $\leqslant$ 0.40.

The data points with large PF, i.e. PF $\approx$ 0.30, represent tightly packed carbon
structures which have predominantly $sp^3$ bonds, whereas data points with PF $\approx$
0.10 represent loosely packed structures which have predominantly $sp^1-sp^2$ bonds.
Datapoints with PF $\approx$ 0.20 are those cases which have intermediate packing
arrangement and consist mainly a mixture of $sp^2-sp^3$ bonds.

Unlike RS database, the MHM database has structures with a wide range of PF, varying composition
of $sp^1-sp^2-sp^3$ and relatively larger $\Delta$E$_{DFT}$. This is because the MHM
database contains intermediate structures which can be tightly, moderately and loosely packed.
And, it is already well known that deviation from the local minimum leads to increase
in energy. The spread in the PF also correlates with the spread in the $d_{CC}$
and E$_{DFT}$ of the RS and MHM structures.


Apart from the influence of the features on the performance of the ML model,
we also wanted to investigate how different databases influence the performance of the
ML model. To this end we created different subsets of the MHM+RS database based on
physically meaningful quantities.

We inspected the trends in data distribution for different combination of SGDs.
Specifically, we observed that the data points seem to cluster into two clusters
when viewed for an arbitrary pair of distinct SGDs. The distribution plots for every
combination of SGD is shown in Fig.~\ref{fig:2SGDdensity} in Appendix~\ref{sec:AppHistogram}.
We found that the combination of bond length $d_{CC}$(\AA) and PF gave the best clustered plot,
i.e. well-separated clusters with minimal overlap. The density plot for these two SGDs is shown
in right lower panel of Fig.~\ref{fig:CarbonDB}. The MHM+RS database breaks down into two clusters:
cluster 1, shown by the region enclosed by the violet circle, represents graphite-like
structures with dominant $sp^2$ bonds, smaller PF and smaller $d_{CC}$, while cluster 2, 
shown by the region enclosed by the green circle, represents diamond-like structures
with dominant $sp^3$ bonds, large PF and large $d_{CC}$. Certain structures of the MHM+RS database
which do not fall in any of the clusters were removed during the ML studies conducted with
cluster 1 and cluster 2. Both cluster 1 and 2 consist of $\sim$5000 structures and, as apparent 
from Fig.~\ref{fig:CarbonDB}, comprise both data points from RS and MHM. Details on the construction 
of cluster 1 and cluster 2 and removal of remaining structures is provided Appendix~\ref{sec:AppMethodology}.

The scatter plot of the data points of cluster 1 (violet squares) and cluster 2 (green circles)
as a function of PF vs $\Delta$E$_{DFT}$ is shown in the right top panel of Fig.~\ref{fig:CarbonDB}.
It is interesting to see that most of the structures in cluster 1 have higher energy i.e. 
$\Delta$E$_{DFT}$ $\geqslant$ 0.6 eV/atom, whereas the majority of structures of cluster 2 
tend to concentrate more in the energy range 0-1 eV/atom. This indicates that structures with
dominant $sp^3$ bonds are energetically more preferable at ambient conditions.

All the ab-initio calculation have been carried out at the level of density functional theory
(DFT) as available in Vienna Ab-initio Simulation Package (VASP)~\cite{VASP_Kresse}. Details of
the DFT calculations, structure generation, methodology for removing the redundant-unphysical
structures, construction of cluster 1 and cluster 2 and removal of remaining structures are
provided in Appendix~\ref{sec:AppMethodology}.

\section{Machine Learning Model: Kernel Ridge Regression}\label{sec:MLmodel}
We model the total energy $E(\textbf{x})$ of carbon crystal structures as a function of our
constructed features $\textbf{x}$, i.e. the function $\textbf{x}\mapsto E(\textbf{x})$, using 
KRR. KRR is a nonparametric regression technique that is capable of
performing complex nonlinear regression by conducting linear regression in an implicit hyperspace.
Due to the so-called kernel trick, no explicit transformation to that hyperspace is needed,
which is why KRR has a great computational advantage over conventional nonlinear regression.
In the KRR formalism, predictions are made according to the similarity between the 
representations of two crystal structures:
\begin{equation}
    E(\bm{x}) = \sum_{i=1}^n\alpha_i \kappa(\bm{x}_i,\bm{x}) \quad,
\end{equation}
where $\kappa$ and $\alpha_i$ describe the similarity function, also referred to as the kernel,
and the kernel weight of structure $i$ respectively. Here, $n$ refers to the number of crystal
structures used to train the algorithm and $\bm{x}_i$ to the feature vector of structure $i$. 

Another great advantage of KRR is the availability of a closed-form solution for the optimal
kernel weights. Specifically, the optimal kernel weights $\bm{\alpha}=(\alpha_1,\dots,\alpha_n)$
are given by
\begin{equation}\label{eq:weights}
    \bm{\alpha} = (K + \sigma I)^{-1}\bm{y} \quad,
\end{equation}
where $\bm{y}=(E(\bm{x}_1),\dots,E(\bm{x}_n))$ is the vector of energies of the training structures
and $K=[\kappa(\bm{x}_i,\bm{x}_j)]_{i,j=1,\dots,n}$ denotes the kernel matrix, which gives the
instance-based similarity between each pair of training structures. The term $\sigma I$ results
from the quadratic loss function of the algorithm, which uses regularization of the model weights
to avoid overfitting. The regularization parameter $\sigma$ has to be optimized prior to making
predictions.

In this paper we have used a radial basis function, or Gaussian kernel, given by
\begin{equation}\label{eq:RBF}
    \kappa(\bm{x}^\prime,\bm{x}) = \exp( -\gamma \|\bm{x}^\prime - \bm{x} \|^2 ), 
\end{equation}
where $\gamma$ denotes the kernel coefficient and $\|\cdot\|$ denotes the Euclidean distance. 
The kernel function is unity for identical structures and decreases as their distance increases
in the feature space. The kernel coefficient $\gamma$ determines the rate of decrease and has
to be optimized simultaneously with the regularization parameter. It becomes clear that the
definition of the Gaussian kernel in eq.~\eqref{eq:RBF} requires a continuous data 
representation to effectively measure the similarity of two structures. 
Tuning the hyperparameters $\gamma$ and $\sigma$ was performed using 5-fold cross 
validation. All the training, testing and validation of the ML models were done
using the Scikit library available in Python and the details are discussed in
Appendix~\ref{sec:AppMethodology}.

\section{RESULTS}\label{sec:Results}
This section presents the (i) influence of feature selection, (ii) influence of database
selection, and (iii) evaluates the resulting ML model performances against the holdout
set and independent test set consisting of 12 carbon allotropes.

In Sec.~\ref{sec:FEATinfluence} the influence of different combination of features
(SGD, ADF, RDF) is investigated through training and testing on the MHM+RS database. Since,
we observed that combining all features leads to the best performance, we studied the
influence of different database, i.e., cluster 1 and 2 in Sec.~\ref{sec:DBinfluence} using
all available features. Finally, in Sec.~\ref{sec:comparison}, we compare the performance
of the ML models trained in Sec.~\ref{sec:FEATinfluence} and Sec.~\ref{sec:DBinfluence}
against the 12 carbon allotropes in the independent test set. Prior to any model
training/testing, we performed standard feature scaling and optimization procedures
where conducted using 5-fold cross validation.

\subsection{Influence of Features}\label{sec:FEATinfluence}
\begin{figure}
\includegraphics[width=1.0\columnwidth,angle=0]{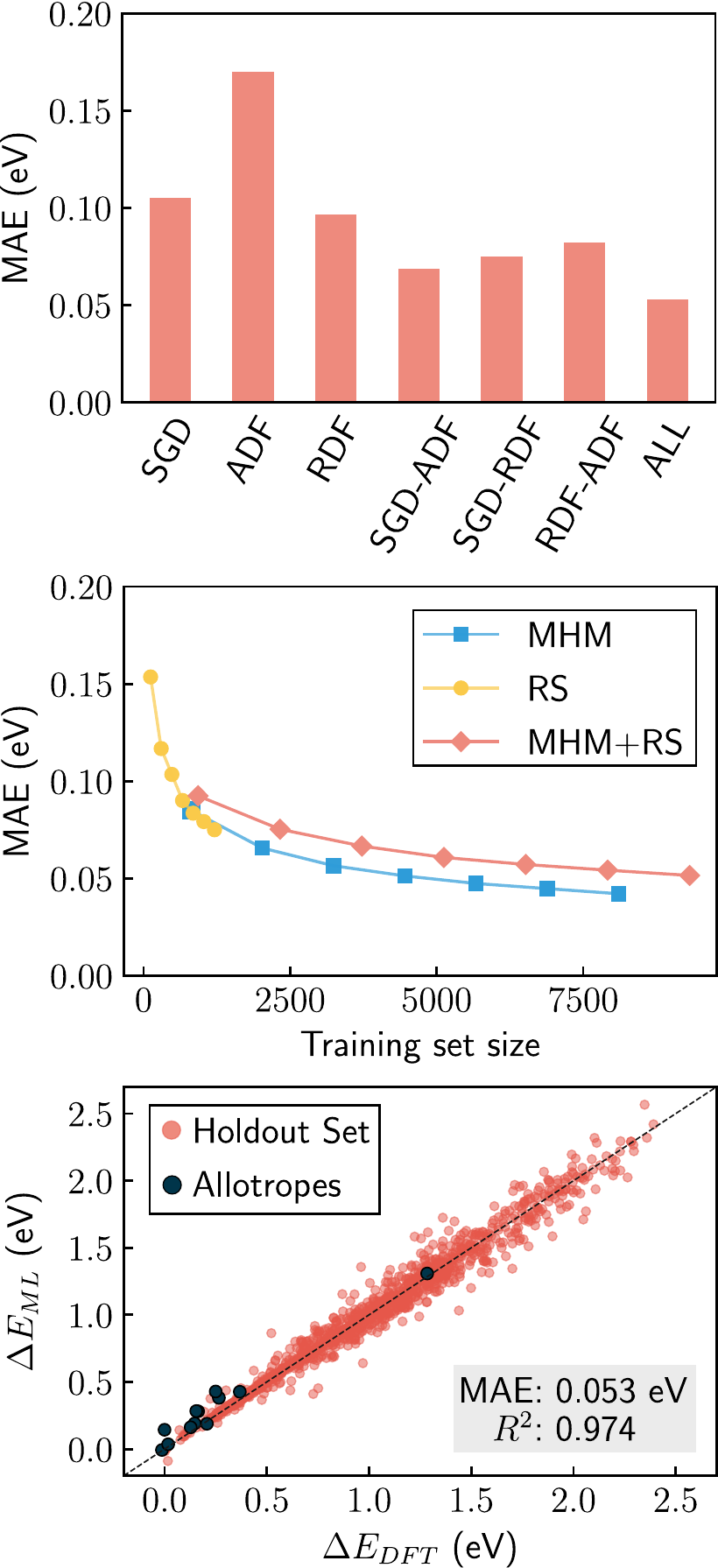}
\caption{
Top panel displays the mean absolute error (MAE) in (eV/atom) of the KRR model trained
using different combination of features (SGD, ADF and RDF) for the MHM+RS (red) database.
Middle panel shows the learning curve of the ML model as a function of training set size
for the MHM (blue squares), RS (yellow circles) and and MHM+RS (red diamonds) database.
Bottom panel displays scatter plot of energy obtained from first principles
$\Delta$E$_{DFT}$ in (eV/atom) (abscissa) against energy predicted by the ML model
$\Delta$E$_{ML}$ (ordinate) trained on the full MHM+RS database with all features. 
The red circles represent the 10 $\%$ holdout set and black dots the 12 carbon
allotropes. The inset lists the MAE and $R^2$ computed on the holdout set.}
\label{fig:FEATinfluence}
\end{figure}

A consistent performance comparison of the KRR model trained with
different combination of features (SGD, RDF, ADF) is only possible if the parameters of the
distribution functions ($\Delta r$, $r_c$, $\Delta \theta$) are kept constant for all tests. 
To this end, we first optimized the parameters of the RDF and ADF on the entire crystal
structure database (MHM+RS) as discussed in Appendix~\ref{sec:AppMethodology}.
The set of optimal parameters are listed in Tab.~\ref{tab:feature}.

Since changing the features implies changing the feature space (and distances within
the feature space), the model parameters ($\gamma$, $\sigma$) have to be optimized for
the corresponding combination of features. Details about the grid search for optimal model
parameters are also provided in Appendix~\ref{sec:AppMethodology}.

One aim of our study was to determine the relative importance of the different feature sets
(SGD, ADF, RDF) introduced in Sec.~\ref{sec:DATA-FEATURES}. Their relative importance becomes
evident in their ML model performance study on MHM+RS database. The top panel of
Fig.~\ref{fig:FEATinfluence} shows the mean absolute error (MAE) found on total energy per
atom (eV/atom) of carbon structures for different subsets of selected features. Note that
the shown mean performance refers to the MAE on the validation sets of the cross validation
scheme and does not necessarily represent the accuracy on a (independent) test set.

The total energy of any condensed matter systems is primarily influenced by inter-atomic distances
of atoms, whereas the angular distribution plays a small but important role. This is clearly
reflected in SGD and RDF features leading to better performance than ADF features alone, as
shown in top panel of Fig.~\ref{fig:FEATinfluence}. It is interesting to note that SGD with
only 8 features is able to achieve similar performance as the 40 computationally relatively
expensive RDF features.

Combining ADF with either RDF or SGD helps to obtain better performance as compared to using
only RDF or SGD features. The performance of different combinations of two types of features
with decreasing MAE is in the following order: (RDF+ADF) $>$ (SGD+RDF) $>$ (SGD+ADF) with MAE 
$\approx$ 70-80 meV/atom. This indicates that a similar performance can be achieved by using SGD+ADF 
with 20 features as compared to RDF+ADF with 52 features. However, the best performance is 
obtained by combining all the three features types, achieving a MAE of 53 meV/atom. 

We also investigated if the method of structure generation influences the results of
model training. For this test we used all (SGD+RDF+ADF) features and trained the KRR model
with different number of training examples chosen from the MHM, RS and MHM+RS database.
For each database, the optimum choice of model parameters was determined in advance,
as discussed in the Appendix~\ref{sec:AppMethodology}.

The learning curves in the middle panel of Fig.~\ref{fig:FEATinfluence} show the MAE 
(mean over 5 folds) of total energy per atom measured on the validation sets.
The learning curves of MHM (blue squares), RS (yellow circles) and MHM+RS (red diamonds)
are similar in nature. With similar kinds of structures in MHM and RS, as evident from
Fig.~\ref{fig:CarbonDB}, their similar learning curves indicate that the performance of
the ML model is mostly influenced by the training set size of the database rather than the method
of generation. The minor increase in the MAE using the MHM+RS database can be explained by an
intermediate choice of optimal hyperparameters (Appendix~\ref{sec:AppMethodology}), which
were neither optimal for the RS database, nor the MHM database.

Finally, the performance of the KRR model trained on the MHM+RS database, represented through
all 60 features with the optimal model parameters is benchmarked against the $10\%$-holdout
set and the test set with the 12 carbon allotropes. 

The predicted energies ($\Delta E_{ML}$), as compared to DFT energies ($\Delta E_{DFT}$),
of the structures of the holdout set and allotropes are shown as solid red and black circles,
respectively, in the bottom plot of Fig.~\ref{fig:FEATinfluence}. The predicted energies of
the structures on the holdout set have a MAE of 53 meV/atom, which corresponds to 97.4 $\%$
in terms of the $R^2$ value, the coefficient of determination. As evident from the plot,
the predicted energies of the 12 carbon allotropes are in good agreement with their DFT
energy. The prediction error on each carbon allotrope in specific is shown and discussed
in Sec.~\ref{sec:comparison}.

\subsection{Influence of database }\label{sec:DBinfluence}
\begin{figure}[!htb]
\includegraphics{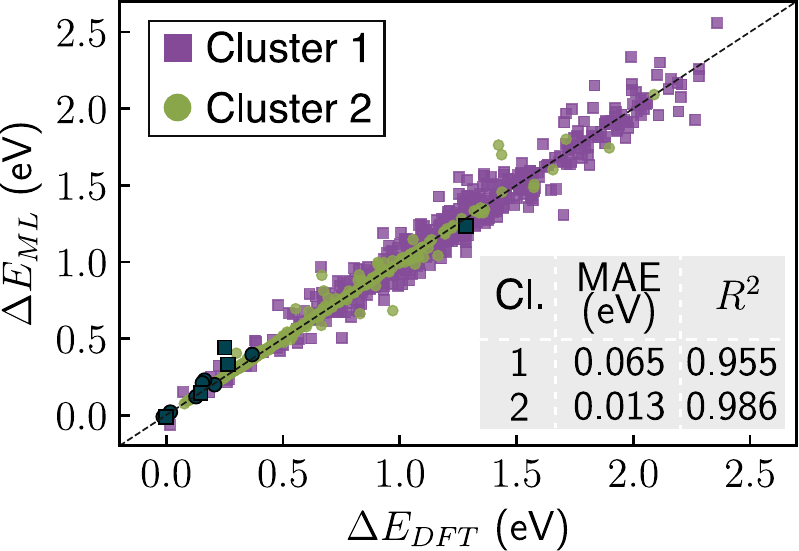}
\caption{Scatter plot of the total energy of the structures in the holdout set and 12 
carbon allotropes obtained from the DFT $\Delta$E$_{DFT}$ vs.\ predicted using the ML model
$\Delta$E$_{ML}$ in eV/atom. The data points of the holdout set and allotropes for cluster 1
are shown by the violet and black squares, respectively, and for cluster 2 by green and black
circles, respectively. The mean absolute errors (MAE) in eV/atom and the R$^2$ values for
cluster 1 and cluster 2 are provided in the inset.}
\label{fig:DBinfluence}
\end{figure}

Along with the influence of features, we were also interested to investigate on how the
model performance is influenced by the different choice of a database. In order to probe this
aspect of ML studies, we created cluster 1 database  consisting of loosely packed graphite-like
structures and cluster 2 database  consisting of tightly packed diamond-like structures as shown
in the lower/upper right panel of Fig.~\ref{fig:CarbonDB}. Similarly, we split the $10\%$-holdout
and test set into the two clusters for later model testing. Detailed information about the
creation and pruning of the these two distinct clusters is provided in Sec.~\ref{sec:CarbonDB}
and Appendix~\ref{sec:AppMethodology}. In order to highlight the influence of the data, in
this section we trained two separate KRR models on cluster 1 database  and cluster 2 database,
respectively, using all 60 features (SGD+RDF+ADF) and the optimum model parameters of the MHM+RS
database as discussed in Sec.~\ref{sec:FEATinfluence}.

The performance of the two ML models on the respective holdout and test sets are shown as
scatter plot of the total energy obtained from DFT, $\Delta$E$_{DFT}$ vs.\ the energy predicted
using the trained ML models $\Delta$E$_{ML}$ in (eV/atom) in Fig.~\ref{fig:DBinfluence}. The
data points of the holdout sets and allotropes in cluster 1 are shown through violet and black
squares, respectively, whereas those of cluster 2 are shown through green and black circles,
respectively. Most of the data points of the holdout set of cluster 2 fall on the proximity
of the dashed line indicating good match. In particular, cluster 2 exhibits a better match
than cluster 1. This is also reflected in the MAE of cluster 1 with 65 meV/atom which is
$\approx$5 times larger than the MAE of cluster 2, which is 13 meV/atom as shown in the
inset of Fig.~\ref{fig:DBinfluence}. The drastic difference in the MAE of cluster 1 and 
cluster 2 clearly shows that the choice of database can have a profound impact on the performance
of the ML model. In particular, it appears that while a ML model can achieve good
performance relatively easily on a specific database, it is the generalization to
a wide variety of structure types which is problematic. Most of the data points of
the allotropes fall on the dashed line validating a good performance of both trained
models.

\subsection{Model Comparison}\label{sec:comparison}
\begin{figure}[!htb]
\includegraphics{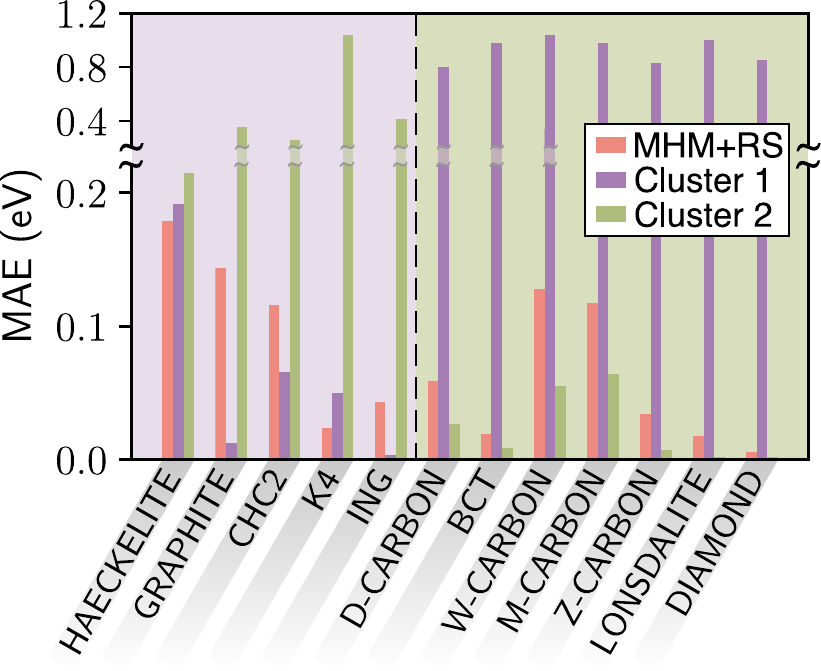}
\caption{Mean absolute error (MAE) in energy of the carbon allotropes predicted by the
trained ML model with MHM+RS (red), Cluster 1 (violet) and Cluster 2 (green) database.}
\label{fig:comparison}
\end{figure}

In the discussions in Sec.~\ref{sec:FEATinfluence} and Sec.~\ref{sec:DBinfluence}, the
performance has been judged through MAE and $R^2$ obtained by the prediction on the holdout
sets, which were selected from the same origin as the model training data. In this section,
we additionally evaluate the performance of three different ML models (trained on the MHM+RS,
cluster 1, and cluster 2 database, respectively) on the 12 carbon allotropes.

In Fig.~\ref{fig:comparison} the MAE on the 12 carbon allotropes is shown in detail.
The structures in this plot are arranged in order of increasing PF. The plot shows that
the compact structures are predicted well by the ML model trained with the cluster 2 database
(green), which fails to yield good results for loose structures. The trend is reversed
for the ML model trained on the cluster 1 database (violet). A curious anamoly is haeckelite 
where the cluster 1 model performs poorly.

From these results it becomes evident that compared to the prediction of the ML model trained
on the MHM+RS database (red) we could improve the model performance on a specific structure type by
just selecting the relevant training data. Equivalently, removing irrelevant structures from
the training data can improve performance.

\section{Conclusions} \label{sec:conclusion}
The primary motivation in this work was to gain insight on how different combinations of
feature sets and database subsets influence the performance of a ML model for predicting a target
property of a condensed matter systems, i.e. the total energy in our case.

In order to carry out the investigation, we created a large pool of carbon crystal
structures through minima hopping method and random search resulting in 11500 distinct
structures in total. We further constructed features of three types: RDF, ADF and single
geometric descriptors (SGD) which possess desirable qualities as discussed in Sec.~\ref{sec:DATA-FEATURES}.
The training and testing of ML models with different combinations of these features and database
led to several interesting observations, which we summarize below:

\begin{enumerate}[(i)]
\item The widely used and popular RDF for representing crystal structures is not sufficient
to achieve desirable accuracy for prediction of polymorphic systems. This is because the
RDF loses vital information due to averaging over atoms and ignoring the dependency of bond
angles, which are crucial for polymorphic system. Combining RDF with either ADF or SGD
help to recover parts of the lost/missing information
and improves the performance of the ML 
model. Especially adding SGD 
results in a feature set that embeds both the local and global information 
of the structure.

\item The simple, human-understandable features in SGD 
not only improve the performance of the ML 
model, but also provide useful insight on
the MHM+RS database. These physically meaningful features made it feasible to clearly
understand what kind of structures one can expect in certain domains of the database.
In our case, the observation of the single geometric descriptor features led to a natural
clustering of the database into graphite-like and diamond-like structures. However, this
observation may not hold true for other systems.

\item The final performance on the holdout set depends not only on the features and the 
ML 
model, but also on the database used for training. As seen in our experiments, one set
of database (cluster 2) achieves very low MAE whereas others have very high MAE with same set
of features. This situation is often encountered in practice and leads to removal of certain
data points for improving performance.
However, this issue had seldom been addressed. Through systematic study, for the first
time we clearly show with proper understanding how different database influence the performance
of the machine learning model. Hence, the performance achieved by a machine learning model is not
universal; rather is always subjective to the set of features and the database used.

\item The true performance of the trained machine learning model can only be judged from the
prediction on the test set, whose data is unseen during model training and hyperparameter selection.
As shown in our tests, none of the considered ML 
models displays good performance against all the 12 carbon
allotropes. Instead, their performances are biased towards certain kinds of structures (e.g.,
diamond- or graphite-like) which are well represented in the database used for training. This
clearly points to the fact that a trained ML model does not necessarily generalize to all
structures in the test set. Rather, it depends to a large extent on the combination of features,
the database used for training, and if the given structures in the test set are well represented
by the database.

\item The ML 
model trained with the features sets
ADF and SGD 
achieve comparable performance to the one with all the feature sets (RDF, ADF, SGD). 
The key difference is that the former requires 20 features whereas the latter requires 
60 features. In a 8 atom unit cell, the overhead computational cost would not be huge. 
However, it becomes significant when a unit cell with a large number of atoms is considered. 
As averaging leads to loss of information in large systems, one has to use a feature vector 
for every atom as the representation for ML. In this situation, our SGD+ADF 
feature set would provide a cheaper and faster alternative as compared to full feature set.

\item The ML 
model trained in our study seems to work well if the right set of
features and database is considered. It can be used for initial screening of tightly packed
carbon structures. But, it would be limited to systems with small number of atoms/unit cell.
This is because the averaging of features over all the atoms in case of structures with large
number of atoms/unit cell would lead to loss of vital information. 
This is clearly seen in our preliminary test of the ML 
model applied on random structures consisting of 60/120/180 atoms per unit cell, resulting in
very high MAE. 

\item The performance of the ML model is not effected by the choice of exchange-correlation
functionals or absence of dispersion correction. Our machine learning studies for the the
Ceperly-Alder-Local Density Approximation (CA-LDA)~\cite{ceperley1980ground} along with the
PBE functional led to similar conclusion as discussed above. Thus,
the conclusions are independent of the exchange-correlation functional.
\end{enumerate}

Hence, our physically motivated SGD were instrumental in seeking an in-depth understanding
of how different factors, i.e. feature sets and database subsets influence the ML model performance.
The performance tests with different combinations of features and database indicate that it is
difficult to construct a general universal ML model which generalizes well to all types of
structures. Instead, one needs to optimize the set of features, database and the ML model for the
target system of interest to obtain the best performance.

\begin{acknowledgements}
F.~M.~Rohrhofer, S. Saha, S. Di Cataldo and W. von der Linden acknowledge computational 
resources from the dCluster of the Graz University of Technology and the VSC3 of the Vienna 
University of Technology, and support through the FWF, Austrian Science Fund, Project
P 30269- N36 (Superhydra).
F.~M.~Rohrhofer and B.~C.~Geiger acknowledge the financial support of the Austrian COMET -
Competence Centers for Excellent Technologies - Programme of the Austrian Federal Ministry
for Climate Action, Environment, Energy, Mobility, Innovation and Technology, the Austrian
Federal Ministry for Digital and Economic Affairs, and the States of Styria, Upper Austria,
Tyrol, and Vienna for the COMET Centers Know-Center and LEC EvoLET, respectively. The COMET
Programme is managed by the Austrian Research Promotion Agency (FFG). L. Boeri acknowledges
support from Fondo Ateneo Sapienza 2017-19 and computational Resources from CINECA, proj.
Hi-TSEPH.
\end{acknowledgements}


\onecolumngrid
\appendix

\section{Methodology}\label{sec:AppMethodology}

\subsection{Data Generation and DFT Calculation}

\subsubsection{Data Generation}
We have used two approaches for generation of dataset of carbon structures for
ML 
studies. In the first approach we have generated structures through RS
approach using the random structure generator as implemented in USPEX
~\cite{glass2006uspex,lyakhov2013new}. In the second approach we have employed MHM
~\cite{goedecker2004minima,amsler2010crystal}, which consists of consecutive short MD runs
followed by  post-relaxation. In the MHM method, one has to provide initial seed structure as
starting point for MHM runs. We have used graphite, diamond, BCT~\cite{liu1992theoretical} and
K4~\cite{sunada2008crystals}
allotropes of carbon as starting point. Structures generated from these two approaches were 
screened for any redundancy~\cite{valle2010crystal} based on fingerprint distance and unphysical
structures based on the PF. This resulted in a final tally of 1500 structures from RS and 10000
from MHM runs. An 8 atom unit cell of carbon was used for the data generation and ML studies.

The fingerprint function developed by Oganov. et al.~\cite{valle2010crystal} based on the 
cosine of the angle formed by the RDF of any two structure was used to sort out the database from
multiple found structures. A well defined mutual distance of $\leqslant$ 0.0001 among 
structures in phase space was set as lower threshold. The unphysical structures represent
those cases, which have low PF, i.e. PF $\leqslant$ 0.05.

\subsubsection{DFT Calculations}
All the calculations in structure generation, geometry relaxation and scf have been carried out
using planewave based density functional theory as available in Vienna Ab-initio Simulation
Package (VASP)~\cite{VASP_Kresse} for Perdew-Burke-Ernzerhof exchange correlation functional
functional~\cite{perdew1996generalized}. An energy cut-off of 400 eV and Gaussian smearing of
0.20 eV was used for quick and efficient generation of structures, without losing on accuracy.

Post data generation through RS and MHM, all the structures of RS and the local minima
structures of MHM were further relaxed to a threshold force of 10 meV/\AA. A higher energy
cutoff of 600 eV, Gaussian smearing of 0.10 eV and a k-mesh of 2$\pi$ x 0.20 \AA$^{-1}$
to sample the Brillouin Zone (BZ) was used for geometric relaxation. In the final step,
these settings were used for all the intermediate and relaxed structures for scf calculations
to obtain final set of consistent total energies.

As our crystal database consists of graphite-like structures, good description of van der
Waal's interaction would be necessary to obtain both correct geometry and correct energy.
However, we have deliberately avoided the use of dispersion correction along with PBE
functional. This decision was motivated by the fact that (i) our constructed features
may not be sensitive to the energy scales of the vdW's interaction and (ii) hence we
wanted to ensure that the ML model learns and provide performance only for the electrons
involved in chemical bonding.

In order to ensure that the performance of the ML model is not effected by the choice
of functionals or absence of dipersion correction, we also carried out the systematic ML
studies on the total energies calculated using Ceperly-Alder-Local Density Approximation (CA-LDA)
functinal~\cite{ceperley1980ground} as available in VASP using the settings discussed above for
PBE functional. 

The carbon atoms were described by the Projector Augmented Wave (PAW) potentials as available
for PBE/CA functionals in VASP~\cite{PAW_Bloch,kresse1999ultrasoft}.

\subsection{Data Representation}

\begin{itemize}
\item
\textbf{Single Geometric Descriptor}: 

As discussed in Sec.~\ref{sec:DATA-FEATURES} the SGD component consists of following
features: the atom's average coordination number ($CN$), average inter-atomic bond
distances ($d_{CC}$), number-density ($ND$) and packing-fraction ($PF$).

In order to ensure continuity and consistency in determining the CN
and $d_{CC}$, we used a self-consistent method by \textit{Limbu}~\cite{limbu2018}.
The method starts with an initial guess of the average bond length $\bar{d}_i$ which is
iteratively updated to the actual average bond length of the $i$-th atom. Using the
interatomic distances $d_{ij}$, the average bond length of the $i$-th atom is updated
according to 
\begin{equation}
	\bar{d}_i = \sum_j d_{ij}p_{ij}, \quad p_{ij}=\frac{e^{f(d_{ij})}}{\sum_j e^{f(d_{ij})}}, \quad f(d_{ij})=\left[1 -\left( \frac{d_{ij}}{\bar{d}_i} \right)^6 \right] \;.
\label{eq:average_bond_length}
\end{equation}
Once the average bond length has converged, the (effective) coordination number of the
$i$-th atom is given by
\begin{equation}
C_i = \sum_j e^{f(d_{ij})} \;.
\end{equation}
The individual coordination numbers $C_i$ and bond lengths $\bar{d}_i$ were averaged to
yield the average coordination number $CN_1$ and bond length $d_{CC, 1}$ of first nearest
neighbours. Additionally we determined the standard deviation for both quantities.
Subsequently, we used the same procedure to determine the coordination number and bond
length of second nearest neighbours, $CN_2$ and $d_{CC, 2}$.

The ND and PF, in general, are intensive quantities, i.e. they
both are independent on the system size, capture the system scale and long-range order.
To ensure that both quantities do not capture redundant information, we determined the
ND under the assumption of rigid atom volumes of equal size and the packing
fraction with individual atom volumes:

\begin{equation}
	ND = \frac{N}{V_\textrm{unit cell}}, \quad
	PF = \frac{\sum_{i=1}^NV_i}{V_{\textrm{unit cell}}} ,
\end{equation}
where $V_i$ denotes the individual atomic volumes, which where obtained by successively
determining the maximum extent each atom can take in its local environment, based on a
hard sphere model of the atom.

\item 
\textbf{Radial Distribution Function}: 
In order to construct the RDF as a finite-size feature vector fulfilling the desired
property of continuity, we quantized ($dr \to \Delta r$) and smoothed eq. (\ref{eq:RDF})
using a Gaussian smoothing on the inter-atomic distances ($d_{ij} \to \mathcal{N}(d_{ij}, \sigma_\text{RDF})$).

For determining the optimum parameters of the RDF, we used the whole crystal structure
database (MHM+RS database, after removing the $10\%$ holdout set) and a grid search scheme
using only the RDF as data representation. In the first test, we constructed a grid of
different values for the bin size $\Delta r \in \{0.01, 0.02, 0.05, 0.10, 0.20, 0.50, 1.00\}$
and equally spaced cutoff radii $r_c \in [1, 10]$ (values given in \AA). Using default model
parameters ($\gamma$=$10^{-3}$ and $\sigma$=$10^{-4}$) for the KRR as implemented in
\textit{Scikit-learn} we performed a 5-fold cross validation in each execution in the grid
search scheme. The best mean squared error (lowest mean over different folds) was achieved
using a bin size of $\Delta r=\SI{0.1}{\angstrom}$ and a cutoff radius of $r_c=\SI{5}{\angstrom}$.
In the second test, we again performed 5-fold cross validation using different choices of
Gaussian smoothing parameters $\sigma_\text{RDF} \in \{0.01, 0.02, 0.05, 0.10, 0.20, 0.50, 1.00\}$.
The best performance was measured at a smoothing parameter equal to the bin size
$\sigma_\text{RDF}=\SI{0.1}{\angstrom}$. The quantization of the RDF with the determined
parameters, hence, gave a final number of 40 bins used as features.

\item 
\textbf{Angular Distribution Function}: Following similar protocols as in RDF,
the distribution of bond angles in the ADF was discretized ($d\theta \to \Delta \theta$)
and smoothed ($\theta_{jik} \to \mathcal{N}(\theta_{jik}, \sigma_\text{ADF})$). 

Again, we conducted a grid search scheme using the MHM+RS database with bin sizes
$\Delta \theta \in \{1, 2, 5, 10, 15, 30\}$ and smoothing
$\sigma_\text{ADF} \in \{1, 2, 5, 10, 15, 30\}$ (values given in $^\circ$).
To provide missing and essential information on the radial distances, we also included
the SGD features during model training. The 5-fold cross validation showed that a bin
size of $\Delta\theta=15^\circ$ and, again, a smoothing equal to the bin size
$\sigma_\text{ADF}=15^\circ$ is optimal for the crystal structures in the database.
With a periodic bond-angle range of $\theta \in [0^\circ$,180$^\circ]$ the quantization
of the ADF resulted in 12 bins.
 
\end{itemize}

\subsection{Model Parameter Optimization}

To determine the optimum value of the regularization parameter $\sigma$ and kernel
coefficient $\gamma$ of the KRR model, we again used 5-fold cross validation in a
grid-search scheme. The grid was constructed using a logarithmic range of $10^0$ to
$10^{-7}$ for both $\sigma$ and $\gamma$. During the search it was ensured that the
optimum lies within the search space.

The model parameter optimization for the model used in Sec.~\ref{sec:FEATinfluence} was
conducted twice: (i) on each combination of features (SGD, RDF, ADF) using only the
MHM+RS database and (ii) on each database (MHM, RS, MHM+RS) using all combination of
features (SGD+RDF+ADF).

\begin{figure}[!htb]
    \includegraphics{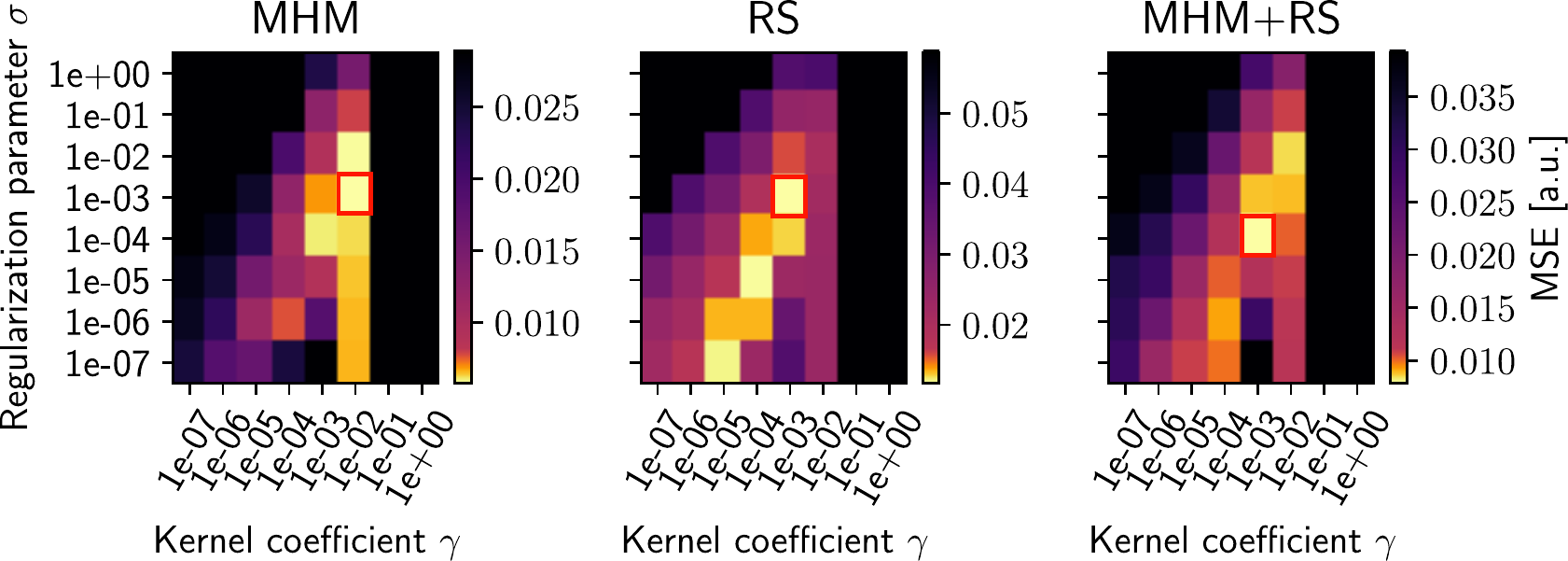}
    \caption{Grid search for the optimal hyperparameters of the KRR model: here, the KRR
    model was trained on each database  (MHM, RS, MHM+RS) using all combination of features
    (SGD+RDF+ADF). The red rectangles indicate the lowest mean squared error in the grid
    search.}
    \label{fig:GRIDsearch}
\end{figure}

A selection of the grid search results for the different databases (MHM, RS, MHM+RS)
is given by Fig.~\ref{fig:GRIDsearch}.

\subsection{Construction of Cluster 1 and Cluster 2}
According to the clusters in Fig.~\ref{fig:CarbonDB}, we identified and labeled two
structure types: loose, $sp^2$-dominated structures (short $d_{CC}$, low PF) as 
\textit{Cluster 1}, and compact and $sp^3$-dominated  structures 
(long bond length, high PF) as \textit{Cluster 2}. 
We performed a subset selection by manually constructing two notional zones, circular and
centered at the cluster peaks, which are apparent in the bottom right panel of Fig.~\ref{fig:CarbonDB}.
The zones were expanded, such that approximately 5000 structures fell into each zone
(Fig.~\ref{fig:CarbonDB}). 
Crystal structures not captured by any of the two zones, were considered to be
not relevant for this specific test. Additionally, we separated the holdout set and test
set by allocating the structures to one of the two structure types.

\subsection{Scripts for Machine Learning Studies}
The extraction and construction of features for every crystal structure of the database,
training and optimization of the model parameters of the KRR model have been carried
out in python and the use of the inbuilt \textit{Scikit-learn} library. 

Extracting the features (RDF, ADF, SGD) for the whole database  (MHM+RS) with 11.5k crystal
structures took $\sim$ 15min on a single processor. The model training on 10k crystal
structures took $\sim$ 20 seconds and a few seconds for the prediction on 1k structures.
In comparison, a single scf calculation for the total energy in DFT requires about
$\sim$ 5-10 CPU minutes with 4 MPI parallel processes. 

\section{Distribution Plots}\label{sec:AppHistogram}

\subsection{Feature Uniqueness}
\begin{figure}[!htb]
    \includegraphics{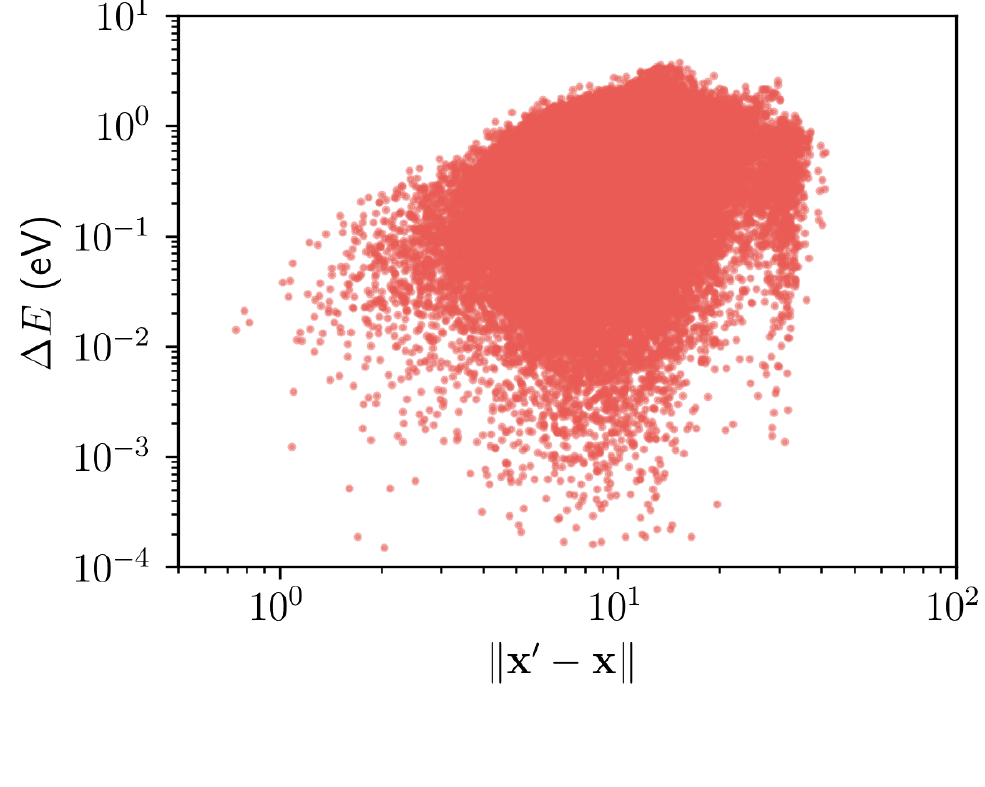}
    \caption{Energy difference $\Delta E$ in (eV/atom) along y-axis vs Euclidean distance
    of the normalized features made of SGD, ADF, RDF along x-axis for every pair of structure
    present in the MHM+RS database.}
    \label{fig:FEATunique}
\end{figure}

In order to check that the features used to represent the carbon structures in our ML
studies are unique, we estimated the Euclidean distances between the set of normalized
features for any two given structures, as they would appear in the definition of the
kernel function eq.(\ref{eq:RBF}). Two distinct structures should result in non-zero
Euclidean distance whereas zero Euclidean distance for two different structure would
indicate that the features are not unique as claimed. A plot of the Euclidean distance
between the normalized features and the energy difference $\Delta$E in (eV/atom) for
every pair of structure is shown in Fig.~\ref{fig:FEATunique}.

\clearpage 

\subsection{Histogram of SGD for RS, MHM and MHM+RS database}
\begin{figure}[!htb]
\includegraphics{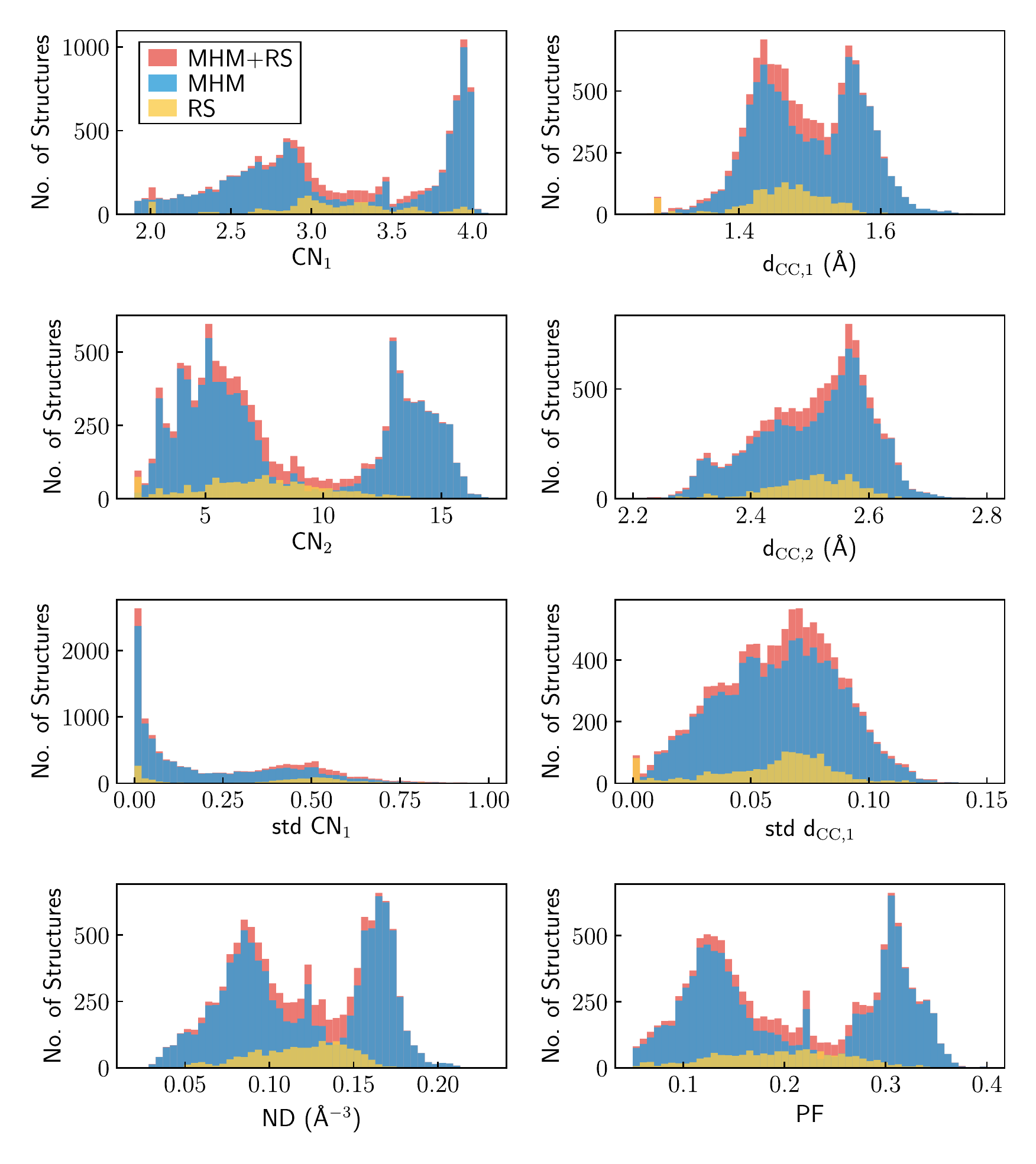}
    \caption{Distribution of structures of MHM(blue), RS(yellow) and MHM+RS(red) for
    different SGD features}
    \label{fig:histfeature}
\end{figure}

\subsection{Histogram of SGD for Cluster 1 and Cluster 2}
\begin{figure}[!htb]
    \includegraphics{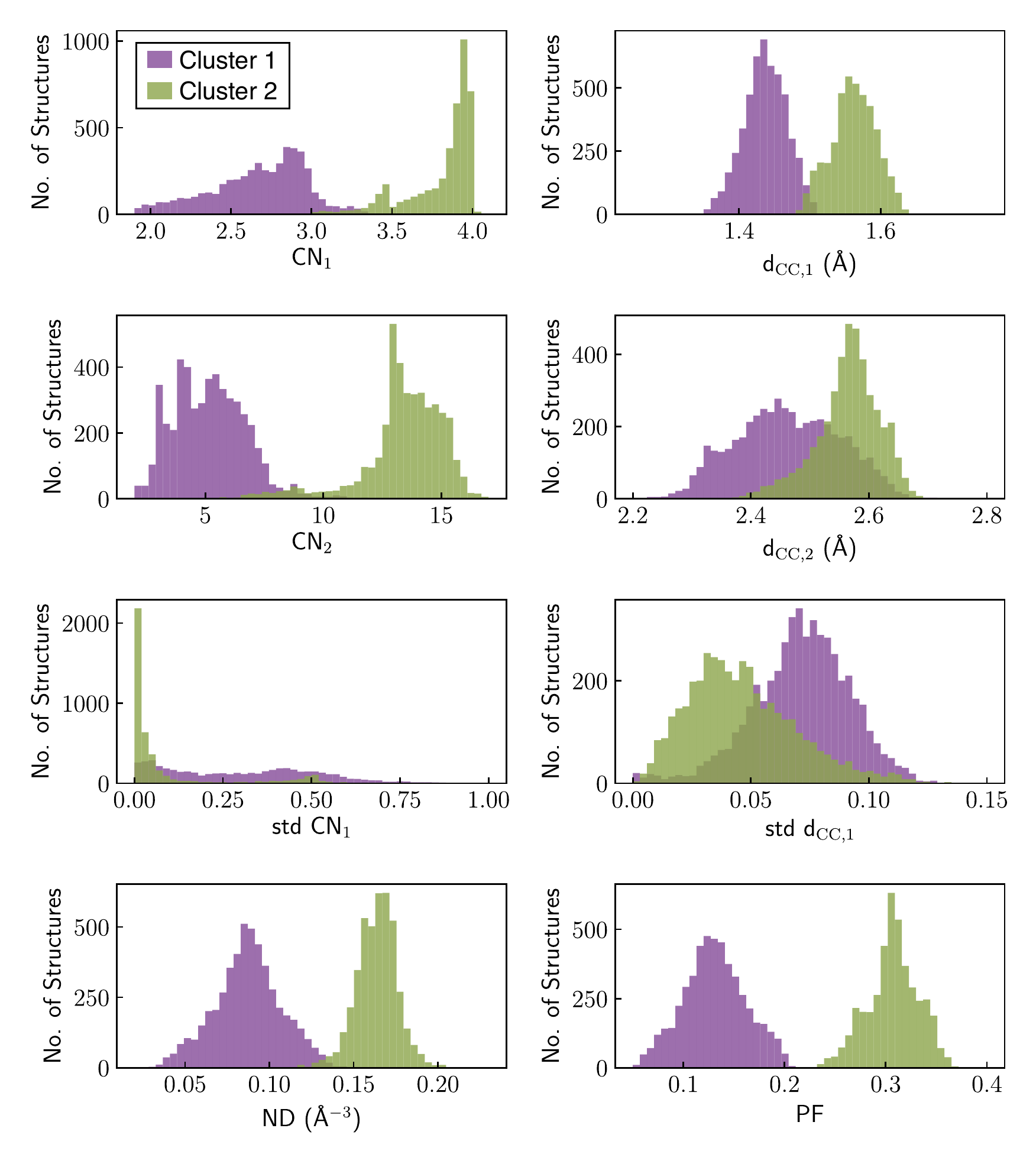}
    \caption{Distribution of structures of Cluster I(violet) and Cluster II(green)
    for different SGD features.}
    \label{fig:histfeaturecluster}
\end{figure}

\subsection{Histogram of RDF and ADF for MHM+RS database}
\begin{figure}[!htb]
    \includegraphics{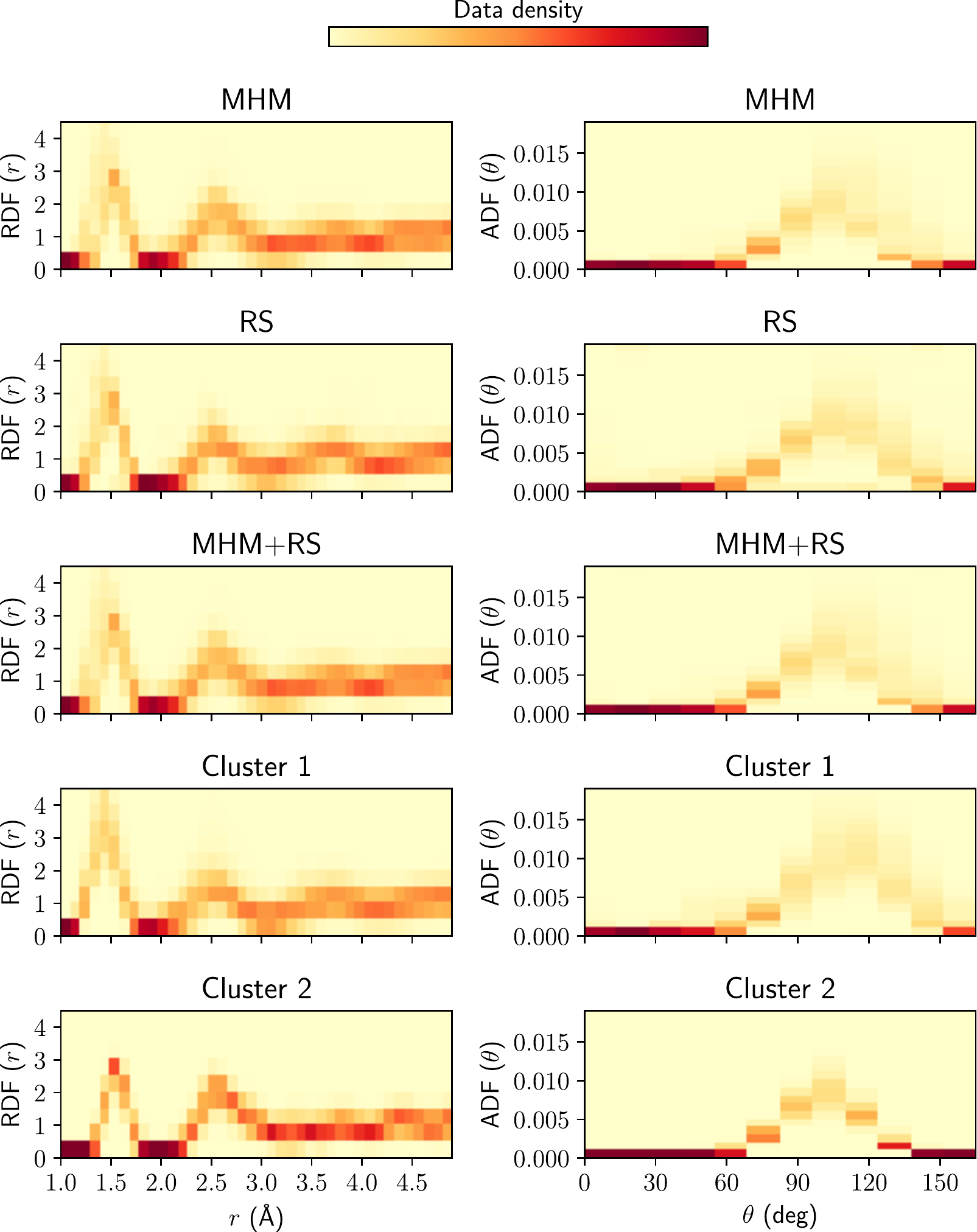}
    \caption{Contour plot of distribution of structures of RDF vs r(\AA)
    and ADF vs $\theta$ for MHM+RS database, Cluster 1 and Cluster 2.
    The color code represents the number of structures.}
    \label{fig:histADFRDF}
\end{figure}

\subsection{Pairwise density plot for different SGD}
\begin{figure}[!htb]
    \includegraphics[width=1\textwidth]{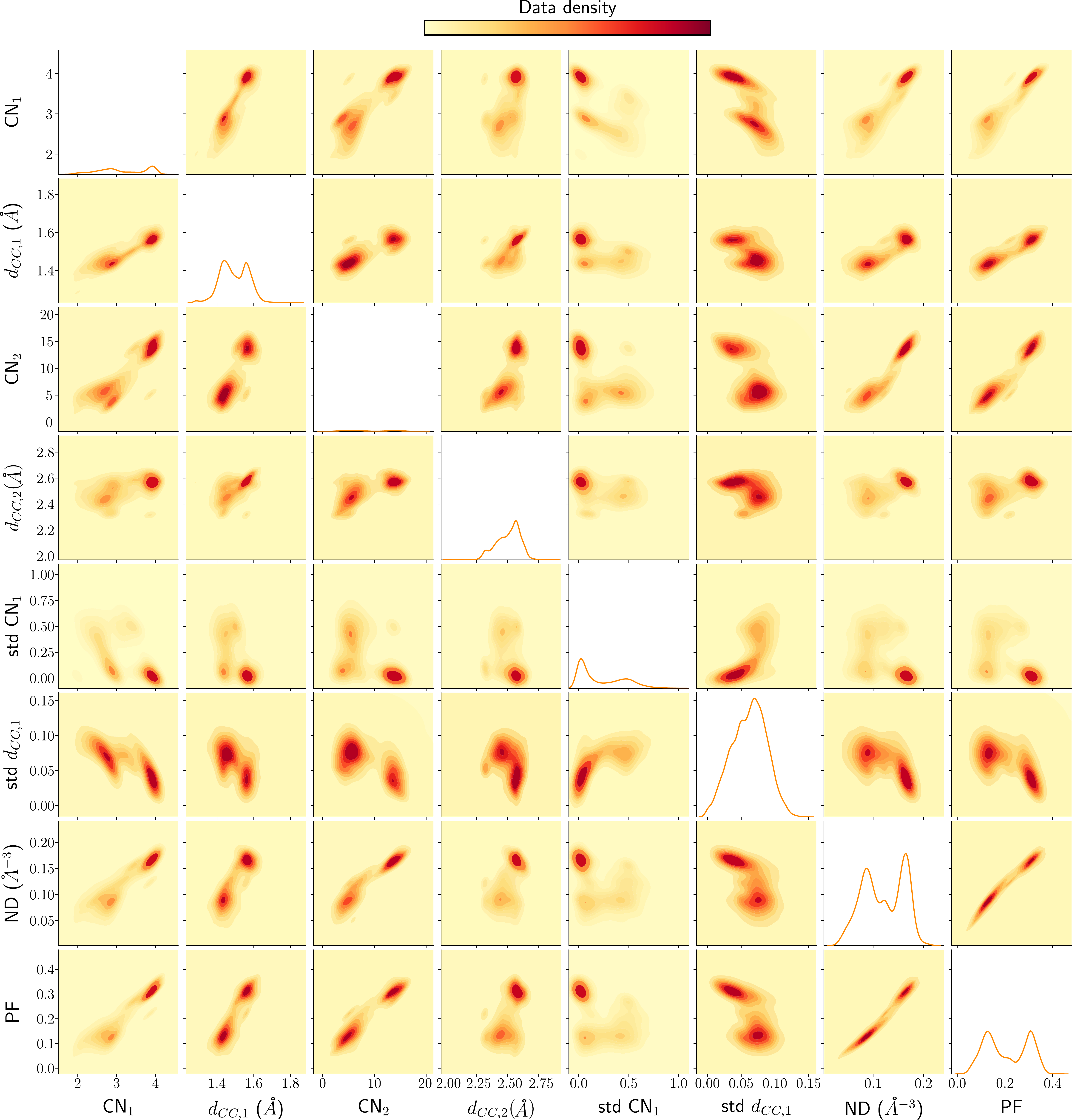}
    \caption{Data density plot for different combination of SGD feature with 
    MHM+RS database.}
    \label{fig:2SGDdensity}
\end{figure}

\twocolumngrid 
\bibliographystyle{apsrev4-1}
\bibliography{main}

\begin{thebibliography}{41}%
\makeatletter
\providecommand \@ifxundefined [1]{%
 \@ifx{#1\undefined}
}%
\providecommand \@ifnum [1]{%
 \ifnum #1\expandafter \@firstoftwo
 \else \expandafter \@secondoftwo
 \fi
}%
\providecommand \@ifx [1]{%
 \ifx #1\expandafter \@firstoftwo
 \else \expandafter \@secondoftwo
 \fi
}%
\providecommand \natexlab [1]{#1}%
\providecommand \enquote  [1]{``#1''}%
\providecommand \bibnamefont  [1]{#1}%
\providecommand \bibfnamefont [1]{#1}%
\providecommand \citenamefont [1]{#1}%
\providecommand \href@noop [0]{\@secondoftwo}%
\providecommand \href [0]{\begingroup \@sanitize@url \@href}%
\providecommand \@href[1]{\@@startlink{#1}\@@href}%
\providecommand \@@href[1]{\endgroup#1\@@endlink}%
\providecommand \@sanitize@url [0]{\catcode `\\12\catcode `\$12\catcode
  `\&12\catcode `\#12\catcode `\^12\catcode `\_12\catcode `\%12\relax}%
\providecommand \@@startlink[1]{}%
\providecommand \@@endlink[0]{}%
\providecommand \url  [0]{\begingroup\@sanitize@url \@url }%
\providecommand \@url [1]{\endgroup\@href {#1}{\urlprefix }}%
\providecommand \urlprefix  [0]{URL }%
\providecommand \Eprint [0]{\href }%
\providecommand \doibase [0]{http://dx.doi.org/}%
\providecommand \selectlanguage [0]{\@gobble}%
\providecommand \bibinfo  [0]{\@secondoftwo}%
\providecommand \bibfield  [0]{\@secondoftwo}%
\providecommand \translation [1]{[#1]}%
\providecommand \BibitemOpen [0]{}%
\providecommand \bibitemStop [0]{}%
\providecommand \bibitemNoStop [0]{.\EOS\space}%
\providecommand \EOS [0]{\spacefactor3000\relax}%
\providecommand \BibitemShut  [1]{\csname bibitem#1\endcsname}%
\let\auto@bib@innerbib\@empty
\bibitem [{\citenamefont {Schleder}\ \emph {et~al.}(2019)\citenamefont
  {Schleder}, \citenamefont {Padilha}, \citenamefont {Acosta}, \citenamefont
  {Costa},\ and\ \citenamefont {Fazzio}}]{schleder2019dft}%
  \BibitemOpen
  \bibfield  {author} {\bibinfo {author} {\bibfnamefont {G.~R.}\ \bibnamefont
  {Schleder}}, \bibinfo {author} {\bibfnamefont {A.~C.}\ \bibnamefont
  {Padilha}}, \bibinfo {author} {\bibfnamefont {C.~M.}\ \bibnamefont {Acosta}},
  \bibinfo {author} {\bibfnamefont {M.}~\bibnamefont {Costa}}, \ and\ \bibinfo
  {author} {\bibfnamefont {A.}~\bibnamefont {Fazzio}},\ }\href@noop {}
  {\bibfield  {journal} {\bibinfo  {journal} {Journal of Physics: Materials}\
  }\textbf {\bibinfo {volume} {2}},\ \bibinfo {pages} {032001} (\bibinfo {year}
  {2019})}\BibitemShut {NoStop}%
\bibitem [{\citenamefont {Lo}\ \emph {et~al.}(2018)\citenamefont {Lo},
  \citenamefont {Rensi}, \citenamefont {Torng},\ and\ \citenamefont
  {Altman}}]{lo2018machine}%
  \BibitemOpen
  \bibfield  {author} {\bibinfo {author} {\bibfnamefont {Y.-C.}\ \bibnamefont
  {Lo}}, \bibinfo {author} {\bibfnamefont {S.~E.}\ \bibnamefont {Rensi}},
  \bibinfo {author} {\bibfnamefont {W.}~\bibnamefont {Torng}}, \ and\ \bibinfo
  {author} {\bibfnamefont {R.~B.}\ \bibnamefont {Altman}},\ }\href@noop {}
  {\bibfield  {journal} {\bibinfo  {journal} {Drug discovery today}\ }\textbf
  {\bibinfo {volume} {23}},\ \bibinfo {pages} {1538} (\bibinfo {year}
  {2018})}\BibitemShut {NoStop}%
\bibitem [{\citenamefont {Kitchin}(2018)}]{kitchin2018machine}%
  \BibitemOpen
  \bibfield  {author} {\bibinfo {author} {\bibfnamefont {J.~R.}\ \bibnamefont
  {Kitchin}},\ }\href@noop {} {\bibfield  {journal} {\bibinfo  {journal}
  {Nature Catalysis}\ }\textbf {\bibinfo {volume} {1}},\ \bibinfo {pages} {230}
  (\bibinfo {year} {2018})}\BibitemShut {NoStop}%
\bibitem [{\citenamefont {Sahu}\ \emph {et~al.}(2018)\citenamefont {Sahu},
  \citenamefont {Rao}, \citenamefont {Troisi},\ and\ \citenamefont
  {Ma}}]{sahu2018toward}%
  \BibitemOpen
  \bibfield  {author} {\bibinfo {author} {\bibfnamefont {H.}~\bibnamefont
  {Sahu}}, \bibinfo {author} {\bibfnamefont {W.}~\bibnamefont {Rao}}, \bibinfo
  {author} {\bibfnamefont {A.}~\bibnamefont {Troisi}}, \ and\ \bibinfo {author}
  {\bibfnamefont {H.}~\bibnamefont {Ma}},\ }\href@noop {} {\bibfield  {journal}
  {\bibinfo  {journal} {Advanced Energy Materials}\ }\textbf {\bibinfo {volume}
  {8}},\ \bibinfo {pages} {1801032} (\bibinfo {year} {2018})}\BibitemShut
  {NoStop}%
\bibitem [{\citenamefont {Haghighatlari}\ \emph {et~al.}(2020)\citenamefont
  {Haghighatlari}, \citenamefont {Li}, \citenamefont {Heidar-Zadeh},
  \citenamefont {Liu}, \citenamefont {Guan},\ and\ \citenamefont
  {Head-Gordon}}]{haghighatlari2020learning}%
  \BibitemOpen
  \bibfield  {author} {\bibinfo {author} {\bibfnamefont {M.}~\bibnamefont
  {Haghighatlari}}, \bibinfo {author} {\bibfnamefont {J.}~\bibnamefont {Li}},
  \bibinfo {author} {\bibfnamefont {F.}~\bibnamefont {Heidar-Zadeh}}, \bibinfo
  {author} {\bibfnamefont {Y.}~\bibnamefont {Liu}}, \bibinfo {author}
  {\bibfnamefont {X.}~\bibnamefont {Guan}}, \ and\ \bibinfo {author}
  {\bibfnamefont {T.}~\bibnamefont {Head-Gordon}},\ }\href@noop {} {\bibfield
  {journal} {\bibinfo  {journal} {Chem}\ } (\bibinfo {year}
  {2020})}\BibitemShut {NoStop}%
\bibitem [{\citenamefont {Behler}(2011)}]{behler2011atom}%
  \BibitemOpen
  \bibfield  {author} {\bibinfo {author} {\bibfnamefont {J.}~\bibnamefont
  {Behler}},\ }\href@noop {} {\bibfield  {journal} {\bibinfo  {journal} {The
  Journal of chemical physics}\ }\textbf {\bibinfo {volume} {134}},\ \bibinfo
  {pages} {074106} (\bibinfo {year} {2011})}\BibitemShut {NoStop}%
\bibitem [{\citenamefont {Smith}\ \emph {et~al.}(2017)\citenamefont {Smith},
  \citenamefont {Isayev},\ and\ \citenamefont {Roitberg}}]{smith2017ani}%
  \BibitemOpen
  \bibfield  {author} {\bibinfo {author} {\bibfnamefont {J.~S.}\ \bibnamefont
  {Smith}}, \bibinfo {author} {\bibfnamefont {O.}~\bibnamefont {Isayev}}, \
  and\ \bibinfo {author} {\bibfnamefont {A.~E.}\ \bibnamefont {Roitberg}},\
  }\href@noop {} {\bibfield  {journal} {\bibinfo  {journal} {Chemical science}\
  }\textbf {\bibinfo {volume} {8}},\ \bibinfo {pages} {3192} (\bibinfo {year}
  {2017})}\BibitemShut {NoStop}%
\bibitem [{\citenamefont {Faber}\ \emph {et~al.}(2018)\citenamefont {Faber},
  \citenamefont {Christensen}, \citenamefont {Huang},\ and\ \citenamefont
  {Von~Lilienfeld}}]{faber2018alchemical}%
  \BibitemOpen
  \bibfield  {author} {\bibinfo {author} {\bibfnamefont {F.~A.}\ \bibnamefont
  {Faber}}, \bibinfo {author} {\bibfnamefont {A.~S.}\ \bibnamefont
  {Christensen}}, \bibinfo {author} {\bibfnamefont {B.}~\bibnamefont {Huang}},
  \ and\ \bibinfo {author} {\bibfnamefont {O.~A.}\ \bibnamefont
  {Von~Lilienfeld}},\ }\href@noop {} {\bibfield  {journal} {\bibinfo  {journal}
  {The Journal of Chemical Physics}\ }\textbf {\bibinfo {volume} {148}},\
  \bibinfo {pages} {241717} (\bibinfo {year} {2018})}\BibitemShut {NoStop}%
\bibitem [{\citenamefont {Christensen}\ \emph {et~al.}(2020)\citenamefont
  {Christensen}, \citenamefont {Bratholm}, \citenamefont {Faber},\ and\
  \citenamefont {Anatole~von Lilienfeld}}]{christensen2020fchl}%
  \BibitemOpen
  \bibfield  {author} {\bibinfo {author} {\bibfnamefont {A.~S.}\ \bibnamefont
  {Christensen}}, \bibinfo {author} {\bibfnamefont {L.~A.}\ \bibnamefont
  {Bratholm}}, \bibinfo {author} {\bibfnamefont {F.~A.}\ \bibnamefont {Faber}},
  \ and\ \bibinfo {author} {\bibfnamefont {O.}~\bibnamefont {Anatole~von
  Lilienfeld}},\ }\href@noop {} {\bibfield  {journal} {\bibinfo  {journal} {The
  Journal of Chemical Physics}\ }\textbf {\bibinfo {volume} {152}},\ \bibinfo
  {pages} {044107} (\bibinfo {year} {2020})}\BibitemShut {NoStop}%
\bibitem [{\citenamefont {Bart{\'o}k}\ \emph {et~al.}(2013)\citenamefont
  {Bart{\'o}k}, \citenamefont {Kondor},\ and\ \citenamefont
  {Cs{\'a}nyi}}]{bartok2013representing}%
  \BibitemOpen
  \bibfield  {author} {\bibinfo {author} {\bibfnamefont {A.~P.}\ \bibnamefont
  {Bart{\'o}k}}, \bibinfo {author} {\bibfnamefont {R.}~\bibnamefont {Kondor}},
  \ and\ \bibinfo {author} {\bibfnamefont {G.}~\bibnamefont {Cs{\'a}nyi}},\
  }\href@noop {} {\bibfield  {journal} {\bibinfo  {journal} {Physical Review
  B}\ }\textbf {\bibinfo {volume} {87}},\ \bibinfo {pages} {184115} (\bibinfo
  {year} {2013})}\BibitemShut {NoStop}%
\bibitem [{\citenamefont {Zhu}\ \emph {et~al.}(2016)\citenamefont {Zhu},
  \citenamefont {Amsler}, \citenamefont {Fuhrer}, \citenamefont {Schaefer},
  \citenamefont {Faraji}, \citenamefont {Rostami}, \citenamefont {Ghasemi},
  \citenamefont {Sadeghi}, \citenamefont {Grauzinyte}, \citenamefont
  {Wolverton} \emph {et~al.}}]{zhu2016fingerprint}%
  \BibitemOpen
  \bibfield  {author} {\bibinfo {author} {\bibfnamefont {L.}~\bibnamefont
  {Zhu}}, \bibinfo {author} {\bibfnamefont {M.}~\bibnamefont {Amsler}},
  \bibinfo {author} {\bibfnamefont {T.}~\bibnamefont {Fuhrer}}, \bibinfo
  {author} {\bibfnamefont {B.}~\bibnamefont {Schaefer}}, \bibinfo {author}
  {\bibfnamefont {S.}~\bibnamefont {Faraji}}, \bibinfo {author} {\bibfnamefont
  {S.}~\bibnamefont {Rostami}}, \bibinfo {author} {\bibfnamefont {S.~A.}\
  \bibnamefont {Ghasemi}}, \bibinfo {author} {\bibfnamefont {A.}~\bibnamefont
  {Sadeghi}}, \bibinfo {author} {\bibfnamefont {M.}~\bibnamefont {Grauzinyte}},
  \bibinfo {author} {\bibfnamefont {C.}~\bibnamefont {Wolverton}},  \emph
  {et~al.},\ }\href@noop {} {\bibfield  {journal} {\bibinfo  {journal} {The
  Journal of chemical physics}\ }\textbf {\bibinfo {volume} {144}},\ \bibinfo
  {pages} {034203} (\bibinfo {year} {2016})}\BibitemShut {NoStop}%
\bibitem [{\citenamefont {Parsaeifard}\ \emph {et~al.}(2020)\citenamefont
  {Parsaeifard}, \citenamefont {De}, \citenamefont {Christensen}, \citenamefont
  {Faber}, \citenamefont {Kocer}, \citenamefont {De}, \citenamefont {Behler},
  \citenamefont {von Lilienfeld},\ and\ \citenamefont
  {Goedecker}}]{parsaeifard2020assessment}%
  \BibitemOpen
  \bibfield  {author} {\bibinfo {author} {\bibfnamefont {B.}~\bibnamefont
  {Parsaeifard}}, \bibinfo {author} {\bibfnamefont {D.~S.}\ \bibnamefont {De}},
  \bibinfo {author} {\bibfnamefont {A.~S.}\ \bibnamefont {Christensen}},
  \bibinfo {author} {\bibfnamefont {F.~A.}\ \bibnamefont {Faber}}, \bibinfo
  {author} {\bibfnamefont {E.}~\bibnamefont {Kocer}}, \bibinfo {author}
  {\bibfnamefont {S.}~\bibnamefont {De}}, \bibinfo {author} {\bibfnamefont
  {J.}~\bibnamefont {Behler}}, \bibinfo {author} {\bibfnamefont
  {A.}~\bibnamefont {von Lilienfeld}}, \ and\ \bibinfo {author} {\bibfnamefont
  {S.}~\bibnamefont {Goedecker}},\ }\href@noop {} {\bibfield  {journal}
  {\bibinfo  {journal} {Machine Learning: Science and Technology}\ } (\bibinfo
  {year} {2020})}\BibitemShut {NoStop}%
\bibitem [{\citenamefont {Rupp}\ \emph {et~al.}(2012)\citenamefont {Rupp},
  \citenamefont {Tkatchenko}, \citenamefont {M{\"u}ller},\ and\ \citenamefont
  {Von~Lilienfeld}}]{rupp2012fast}%
  \BibitemOpen
  \bibfield  {author} {\bibinfo {author} {\bibfnamefont {M.}~\bibnamefont
  {Rupp}}, \bibinfo {author} {\bibfnamefont {A.}~\bibnamefont {Tkatchenko}},
  \bibinfo {author} {\bibfnamefont {K.-R.}\ \bibnamefont {M{\"u}ller}}, \ and\
  \bibinfo {author} {\bibfnamefont {O.~A.}\ \bibnamefont {Von~Lilienfeld}},\
  }\href@noop {} {\bibfield  {journal} {\bibinfo  {journal} {Physical review
  letters}\ }\textbf {\bibinfo {volume} {108}},\ \bibinfo {pages} {058301}
  (\bibinfo {year} {2012})}\BibitemShut {NoStop}%
\bibitem [{\citenamefont {Hansen}\ \emph {et~al.}(2013)\citenamefont {Hansen},
  \citenamefont {Montavon}, \citenamefont {Biegler}, \citenamefont {Fazli},
  \citenamefont {Rupp}, \citenamefont {Scheffler}, \citenamefont
  {Von~Lilienfeld}, \citenamefont {Tkatchenko},\ and\ \citenamefont
  {Muller}}]{hansen2013assessment}%
  \BibitemOpen
  \bibfield  {author} {\bibinfo {author} {\bibfnamefont {K.}~\bibnamefont
  {Hansen}}, \bibinfo {author} {\bibfnamefont {G.}~\bibnamefont {Montavon}},
  \bibinfo {author} {\bibfnamefont {F.}~\bibnamefont {Biegler}}, \bibinfo
  {author} {\bibfnamefont {S.}~\bibnamefont {Fazli}}, \bibinfo {author}
  {\bibfnamefont {M.}~\bibnamefont {Rupp}}, \bibinfo {author} {\bibfnamefont
  {M.}~\bibnamefont {Scheffler}}, \bibinfo {author} {\bibfnamefont {O.~A.}\
  \bibnamefont {Von~Lilienfeld}}, \bibinfo {author} {\bibfnamefont
  {A.}~\bibnamefont {Tkatchenko}}, \ and\ \bibinfo {author} {\bibfnamefont
  {K.-R.}\ \bibnamefont {Muller}},\ }\href@noop {} {\bibfield  {journal}
  {\bibinfo  {journal} {Journal of Chemical Theory and Computation}\ }\textbf
  {\bibinfo {volume} {9}},\ \bibinfo {pages} {3404} (\bibinfo {year}
  {2013})}\BibitemShut {NoStop}%
\bibitem [{\citenamefont {Montavon}\ \emph {et~al.}(2013)\citenamefont
  {Montavon}, \citenamefont {Rupp}, \citenamefont {Gobre}, \citenamefont
  {Vazquez-Mayagoitia}, \citenamefont {Hansen}, \citenamefont {Tkatchenko},
  \citenamefont {M{\"u}ller},\ and\ \citenamefont
  {Von~Lilienfeld}}]{montavon2013machine}%
  \BibitemOpen
  \bibfield  {author} {\bibinfo {author} {\bibfnamefont {G.}~\bibnamefont
  {Montavon}}, \bibinfo {author} {\bibfnamefont {M.}~\bibnamefont {Rupp}},
  \bibinfo {author} {\bibfnamefont {V.}~\bibnamefont {Gobre}}, \bibinfo
  {author} {\bibfnamefont {A.}~\bibnamefont {Vazquez-Mayagoitia}}, \bibinfo
  {author} {\bibfnamefont {K.}~\bibnamefont {Hansen}}, \bibinfo {author}
  {\bibfnamefont {A.}~\bibnamefont {Tkatchenko}}, \bibinfo {author}
  {\bibfnamefont {K.-R.}\ \bibnamefont {M{\"u}ller}}, \ and\ \bibinfo {author}
  {\bibfnamefont {O.~A.}\ \bibnamefont {Von~Lilienfeld}},\ }\href@noop {}
  {\bibfield  {journal} {\bibinfo  {journal} {New Journal of Physics}\ }\textbf
  {\bibinfo {volume} {15}},\ \bibinfo {pages} {095003} (\bibinfo {year}
  {2013})}\BibitemShut {NoStop}%
\bibitem [{\citenamefont {Rowe}\ \emph {et~al.}(2020)\citenamefont {Rowe},
  \citenamefont {Deringer}, \citenamefont {Gasparotto}, \citenamefont
  {Cs{\'a}nyi},\ and\ \citenamefont {Michaelides}}]{rowe2020accurate}%
  \BibitemOpen
  \bibfield  {author} {\bibinfo {author} {\bibfnamefont {P.}~\bibnamefont
  {Rowe}}, \bibinfo {author} {\bibfnamefont {V.~L.}\ \bibnamefont {Deringer}},
  \bibinfo {author} {\bibfnamefont {P.}~\bibnamefont {Gasparotto}}, \bibinfo
  {author} {\bibfnamefont {G.}~\bibnamefont {Cs{\'a}nyi}}, \ and\ \bibinfo
  {author} {\bibfnamefont {A.}~\bibnamefont {Michaelides}},\ }\href@noop {}
  {\bibfield  {journal} {\bibinfo  {journal} {The Journal of Chemical Physics}\
  }\textbf {\bibinfo {volume} {153}},\ \bibinfo {pages} {034702} (\bibinfo
  {year} {2020})}\BibitemShut {NoStop}%
\bibitem [{\citenamefont {Deringer}\ \emph {et~al.}(2018)\citenamefont
  {Deringer}, \citenamefont {Pickard},\ and\ \citenamefont
  {Cs{\'a}nyi}}]{deringer2018data}%
  \BibitemOpen
  \bibfield  {author} {\bibinfo {author} {\bibfnamefont {V.~L.}\ \bibnamefont
  {Deringer}}, \bibinfo {author} {\bibfnamefont {C.~J.}\ \bibnamefont
  {Pickard}}, \ and\ \bibinfo {author} {\bibfnamefont {G.}~\bibnamefont
  {Cs{\'a}nyi}},\ }\href@noop {} {\bibfield  {journal} {\bibinfo  {journal}
  {Physical review letters}\ }\textbf {\bibinfo {volume} {120}},\ \bibinfo
  {pages} {156001} (\bibinfo {year} {2018})}\BibitemShut {NoStop}%
\bibitem [{\citenamefont {Engel}\ \emph {et~al.}(2018)\citenamefont {Engel},
  \citenamefont {Anelli}, \citenamefont {Ceriotti}, \citenamefont {Pickard},\
  and\ \citenamefont {Needs}}]{engel2018mapping}%
  \BibitemOpen
  \bibfield  {author} {\bibinfo {author} {\bibfnamefont {E.~A.}\ \bibnamefont
  {Engel}}, \bibinfo {author} {\bibfnamefont {A.}~\bibnamefont {Anelli}},
  \bibinfo {author} {\bibfnamefont {M.}~\bibnamefont {Ceriotti}}, \bibinfo
  {author} {\bibfnamefont {C.~J.}\ \bibnamefont {Pickard}}, \ and\ \bibinfo
  {author} {\bibfnamefont {R.~J.}\ \bibnamefont {Needs}},\ }\href@noop {}
  {\bibfield  {journal} {\bibinfo  {journal} {Nature communications}\ }\textbf
  {\bibinfo {volume} {9}},\ \bibinfo {pages} {1} (\bibinfo {year}
  {2018})}\BibitemShut {NoStop}%
\bibitem [{\citenamefont {Cheng}\ \emph {et~al.}(2020)\citenamefont {Cheng},
  \citenamefont {Mazzola}, \citenamefont {Pickard},\ and\ \citenamefont
  {Ceriotti}}]{cheng2020evidence}%
  \BibitemOpen
  \bibfield  {author} {\bibinfo {author} {\bibfnamefont {B.}~\bibnamefont
  {Cheng}}, \bibinfo {author} {\bibfnamefont {G.}~\bibnamefont {Mazzola}},
  \bibinfo {author} {\bibfnamefont {C.~J.}\ \bibnamefont {Pickard}}, \ and\
  \bibinfo {author} {\bibfnamefont {M.}~\bibnamefont {Ceriotti}},\ }\href@noop
  {} {\bibfield  {journal} {\bibinfo  {journal} {Nature}\ }\textbf {\bibinfo
  {volume} {585}},\ \bibinfo {pages} {217} (\bibinfo {year}
  {2020})}\BibitemShut {NoStop}%
\bibitem [{\citenamefont {Ward}\ \emph {et~al.}(2017)\citenamefont {Ward},
  \citenamefont {Liu}, \citenamefont {Krishna}, \citenamefont {Hegde},
  \citenamefont {Agrawal}, \citenamefont {Choudhary},\ and\ \citenamefont
  {Wolverton}}]{ward2017including}%
  \BibitemOpen
  \bibfield  {author} {\bibinfo {author} {\bibfnamefont {L.}~\bibnamefont
  {Ward}}, \bibinfo {author} {\bibfnamefont {R.}~\bibnamefont {Liu}}, \bibinfo
  {author} {\bibfnamefont {A.}~\bibnamefont {Krishna}}, \bibinfo {author}
  {\bibfnamefont {V.~I.}\ \bibnamefont {Hegde}}, \bibinfo {author}
  {\bibfnamefont {A.}~\bibnamefont {Agrawal}}, \bibinfo {author} {\bibfnamefont
  {A.}~\bibnamefont {Choudhary}}, \ and\ \bibinfo {author} {\bibfnamefont
  {C.}~\bibnamefont {Wolverton}},\ }\href@noop {} {\bibfield  {journal}
  {\bibinfo  {journal} {Physical Review B}\ }\textbf {\bibinfo {volume} {96}},\
  \bibinfo {pages} {024104} (\bibinfo {year} {2017})}\BibitemShut {NoStop}%
\bibitem [{\citenamefont {Honrao}\ \emph {et~al.}(2019)\citenamefont {Honrao},
  \citenamefont {Anthonio}, \citenamefont {Ramanathan}, \citenamefont
  {Gabriel},\ and\ \citenamefont {Hennig}}]{honrao2019machine}%
  \BibitemOpen
  \bibfield  {author} {\bibinfo {author} {\bibfnamefont {S.}~\bibnamefont
  {Honrao}}, \bibinfo {author} {\bibfnamefont {B.~E.}\ \bibnamefont
  {Anthonio}}, \bibinfo {author} {\bibfnamefont {R.}~\bibnamefont
  {Ramanathan}}, \bibinfo {author} {\bibfnamefont {J.~J.}\ \bibnamefont
  {Gabriel}}, \ and\ \bibinfo {author} {\bibfnamefont {R.~G.}\ \bibnamefont
  {Hennig}},\ }\href@noop {} {\bibfield  {journal} {\bibinfo  {journal}
  {Computational Materials Science}\ }\textbf {\bibinfo {volume} {158}},\
  \bibinfo {pages} {414} (\bibinfo {year} {2019})}\BibitemShut {NoStop}%
\bibitem [{\citenamefont {Bergerhoff}\ \emph {et~al.}(1983)\citenamefont
  {Bergerhoff}, \citenamefont {Hundt}, \citenamefont {Sievers},\ and\
  \citenamefont {Brown}}]{bergerhoff1983inorganic}%
  \BibitemOpen
  \bibfield  {author} {\bibinfo {author} {\bibfnamefont {G.}~\bibnamefont
  {Bergerhoff}}, \bibinfo {author} {\bibfnamefont {R.}~\bibnamefont {Hundt}},
  \bibinfo {author} {\bibfnamefont {R.}~\bibnamefont {Sievers}}, \ and\
  \bibinfo {author} {\bibfnamefont {I.}~\bibnamefont {Brown}},\ }\href@noop {}
  {\bibfield  {journal} {\bibinfo  {journal} {Journal of chemical information
  and computer sciences}\ }\textbf {\bibinfo {volume} {23}},\ \bibinfo {pages}
  {66} (\bibinfo {year} {1983})}\BibitemShut {NoStop}%
\bibitem [{\citenamefont {Belsky}\ \emph {et~al.}(2002)\citenamefont {Belsky},
  \citenamefont {Hellenbrandt}, \citenamefont {Karen},\ and\ \citenamefont
  {Luksch}}]{belsky2002new}%
  \BibitemOpen
  \bibfield  {author} {\bibinfo {author} {\bibfnamefont {A.}~\bibnamefont
  {Belsky}}, \bibinfo {author} {\bibfnamefont {M.}~\bibnamefont
  {Hellenbrandt}}, \bibinfo {author} {\bibfnamefont {V.~L.}\ \bibnamefont
  {Karen}}, \ and\ \bibinfo {author} {\bibfnamefont {P.}~\bibnamefont
  {Luksch}},\ }\href@noop {} {\bibfield  {journal} {\bibinfo  {journal} {Acta
  Crystallographica Section B: Structural Science}\ }\textbf {\bibinfo {volume}
  {58}},\ \bibinfo {pages} {364} (\bibinfo {year} {2002})}\BibitemShut
  {NoStop}%
\bibitem [{\citenamefont {Jain}\ \emph {et~al.}(2013)\citenamefont {Jain},
  \citenamefont {Ong}, \citenamefont {Hautier}, \citenamefont {Chen},
  \citenamefont {Richards}, \citenamefont {Dacek}, \citenamefont {Cholia},
  \citenamefont {Gunter}, \citenamefont {Skinner}, \citenamefont {Ceder} \emph
  {et~al.}}]{jain2013commentary}%
  \BibitemOpen
  \bibfield  {author} {\bibinfo {author} {\bibfnamefont {A.}~\bibnamefont
  {Jain}}, \bibinfo {author} {\bibfnamefont {S.~P.}\ \bibnamefont {Ong}},
  \bibinfo {author} {\bibfnamefont {G.}~\bibnamefont {Hautier}}, \bibinfo
  {author} {\bibfnamefont {W.}~\bibnamefont {Chen}}, \bibinfo {author}
  {\bibfnamefont {W.~D.}\ \bibnamefont {Richards}}, \bibinfo {author}
  {\bibfnamefont {S.}~\bibnamefont {Dacek}}, \bibinfo {author} {\bibfnamefont
  {S.}~\bibnamefont {Cholia}}, \bibinfo {author} {\bibfnamefont
  {D.}~\bibnamefont {Gunter}}, \bibinfo {author} {\bibfnamefont
  {D.}~\bibnamefont {Skinner}}, \bibinfo {author} {\bibfnamefont
  {G.}~\bibnamefont {Ceder}},  \emph {et~al.},\ }\href@noop {} {\bibfield
  {journal} {\bibinfo  {journal} {Apl Materials}\ }\textbf {\bibinfo {volume}
  {1}},\ \bibinfo {pages} {011002} (\bibinfo {year} {2013})}\BibitemShut
  {NoStop}%
\bibitem [{\citenamefont {Saal}\ \emph {et~al.}(2013)\citenamefont {Saal},
  \citenamefont {Kirklin}, \citenamefont {Aykol}, \citenamefont {Meredig},\
  and\ \citenamefont {Wolverton}}]{saal2013materials}%
  \BibitemOpen
  \bibfield  {author} {\bibinfo {author} {\bibfnamefont {J.~E.}\ \bibnamefont
  {Saal}}, \bibinfo {author} {\bibfnamefont {S.}~\bibnamefont {Kirklin}},
  \bibinfo {author} {\bibfnamefont {M.}~\bibnamefont {Aykol}}, \bibinfo
  {author} {\bibfnamefont {B.}~\bibnamefont {Meredig}}, \ and\ \bibinfo
  {author} {\bibfnamefont {C.}~\bibnamefont {Wolverton}},\ }\href@noop {}
  {\bibfield  {journal} {\bibinfo  {journal} {Jom}\ }\textbf {\bibinfo {volume}
  {65}},\ \bibinfo {pages} {1501} (\bibinfo {year} {2013})}\BibitemShut
  {NoStop}%
\bibitem [{\citenamefont {Glass}\ \emph {et~al.}(2006)\citenamefont {Glass},
  \citenamefont {Oganov},\ and\ \citenamefont {Hansen}}]{glass2006uspex}%
  \BibitemOpen
  \bibfield  {author} {\bibinfo {author} {\bibfnamefont {C.~W.}\ \bibnamefont
  {Glass}}, \bibinfo {author} {\bibfnamefont {A.~R.}\ \bibnamefont {Oganov}}, \
  and\ \bibinfo {author} {\bibfnamefont {N.}~\bibnamefont {Hansen}},\
  }\href@noop {} {\bibfield  {journal} {\bibinfo  {journal} {Computer physics
  communications}\ }\textbf {\bibinfo {volume} {175}},\ \bibinfo {pages} {713}
  (\bibinfo {year} {2006})}\BibitemShut {NoStop}%
\bibitem [{\citenamefont {Lyakhov}\ \emph {et~al.}(2013)\citenamefont
  {Lyakhov}, \citenamefont {Oganov}, \citenamefont {Stokes},\ and\
  \citenamefont {Zhu}}]{lyakhov2013new}%
  \BibitemOpen
  \bibfield  {author} {\bibinfo {author} {\bibfnamefont {A.~O.}\ \bibnamefont
  {Lyakhov}}, \bibinfo {author} {\bibfnamefont {A.~R.}\ \bibnamefont {Oganov}},
  \bibinfo {author} {\bibfnamefont {H.~T.}\ \bibnamefont {Stokes}}, \ and\
  \bibinfo {author} {\bibfnamefont {Q.}~\bibnamefont {Zhu}},\ }\href@noop {}
  {\bibfield  {journal} {\bibinfo  {journal} {Computer Physics Communications}\
  }\textbf {\bibinfo {volume} {184}},\ \bibinfo {pages} {1172} (\bibinfo {year}
  {2013})}\BibitemShut {NoStop}%
\bibitem [{\citenamefont {Pickard}\ and\ \citenamefont
  {Needs}(2011)}]{pickard2011ab}%
  \BibitemOpen
  \bibfield  {author} {\bibinfo {author} {\bibfnamefont {C.~J.}\ \bibnamefont
  {Pickard}}\ and\ \bibinfo {author} {\bibfnamefont {R.}~\bibnamefont
  {Needs}},\ }\href@noop {} {\bibfield  {journal} {\bibinfo  {journal} {Journal
  of Physics: Condensed Matter}\ }\textbf {\bibinfo {volume} {23}},\ \bibinfo
  {pages} {053201} (\bibinfo {year} {2011})}\BibitemShut {NoStop}%
\bibitem [{\citenamefont {Goedecker}(2004)}]{goedecker2004minima}%
  \BibitemOpen
  \bibfield  {author} {\bibinfo {author} {\bibfnamefont {S.}~\bibnamefont
  {Goedecker}},\ }\href@noop {} {\bibfield  {journal} {\bibinfo  {journal} {The
  Journal of chemical physics}\ }\textbf {\bibinfo {volume} {120}},\ \bibinfo
  {pages} {9911} (\bibinfo {year} {2004})}\BibitemShut {NoStop}%
\bibitem [{\citenamefont {Amsler}\ and\ \citenamefont
  {Goedecker}(2010)}]{amsler2010crystal}%
  \BibitemOpen
  \bibfield  {author} {\bibinfo {author} {\bibfnamefont {M.}~\bibnamefont
  {Amsler}}\ and\ \bibinfo {author} {\bibfnamefont {S.}~\bibnamefont
  {Goedecker}},\ }\href@noop {} {\bibfield  {journal} {\bibinfo  {journal} {The
  Journal of chemical physics}\ }\textbf {\bibinfo {volume} {133}},\ \bibinfo
  {pages} {224104} (\bibinfo {year} {2010})}\BibitemShut {NoStop}%
\bibitem [{\citenamefont {Behler}\ and\ \citenamefont
  {Parrinello}(2007)}]{behler2007generalized}%
  \BibitemOpen
  \bibfield  {author} {\bibinfo {author} {\bibfnamefont {J.}~\bibnamefont
  {Behler}}\ and\ \bibinfo {author} {\bibfnamefont {M.}~\bibnamefont
  {Parrinello}},\ }\href@noop {} {\bibfield  {journal} {\bibinfo  {journal}
  {Physical review letters}\ }\textbf {\bibinfo {volume} {98}},\ \bibinfo
  {pages} {146401} (\bibinfo {year} {2007})}\BibitemShut {NoStop}%
\bibitem [{\citenamefont {Sch{\"u}tt}\ \emph {et~al.}(2014)\citenamefont
  {Sch{\"u}tt}, \citenamefont {Glawe}, \citenamefont {Brockherde},
  \citenamefont {Sanna}, \citenamefont {M{\"u}ller},\ and\ \citenamefont
  {Gross}}]{schutt2014represent}%
  \BibitemOpen
  \bibfield  {author} {\bibinfo {author} {\bibfnamefont {K.~T.}\ \bibnamefont
  {Sch{\"u}tt}}, \bibinfo {author} {\bibfnamefont {H.}~\bibnamefont {Glawe}},
  \bibinfo {author} {\bibfnamefont {F.}~\bibnamefont {Brockherde}}, \bibinfo
  {author} {\bibfnamefont {A.}~\bibnamefont {Sanna}}, \bibinfo {author}
  {\bibfnamefont {K.-R.}\ \bibnamefont {M{\"u}ller}}, \ and\ \bibinfo {author}
  {\bibfnamefont {E.~K.}\ \bibnamefont {Gross}},\ }\href@noop {} {\bibfield
  {journal} {\bibinfo  {journal} {Physical Review B}\ }\textbf {\bibinfo
  {volume} {89}},\ \bibinfo {pages} {205118} (\bibinfo {year}
  {2014})}\BibitemShut {NoStop}%
\bibitem [{\citenamefont {Valle}\ and\ \citenamefont
  {Oganov}(2010)}]{valle2010crystal}%
  \BibitemOpen
  \bibfield  {author} {\bibinfo {author} {\bibfnamefont {M.}~\bibnamefont
  {Valle}}\ and\ \bibinfo {author} {\bibfnamefont {A.~R.}\ \bibnamefont
  {Oganov}},\ }\href@noop {} {\bibfield  {journal} {\bibinfo  {journal} {Acta
  Crystallographica Section A: Foundations of Crystallography}\ }\textbf
  {\bibinfo {volume} {66}},\ \bibinfo {pages} {507} (\bibinfo {year}
  {2010})}\BibitemShut {NoStop}%
\bibitem [{\citenamefont {Kresse}\ and\ \citenamefont
  {J.}(1996)}]{VASP_Kresse}%
  \BibitemOpen
  \bibfield  {author} {\bibinfo {author} {\bibfnamefont {G.}~\bibnamefont
  {Kresse}}\ and\ \bibinfo {author} {\bibfnamefont {F.}~\bibnamefont {J.}},\
  }\href@noop {} {\bibfield  {journal} {\bibinfo  {journal} {Comput. Mat.
  Sci.}\ }\textbf {\bibinfo {volume} {6}},\ \bibinfo {pages} {15} (\bibinfo
  {year} {1996})}\BibitemShut {NoStop}%
\bibitem [{\citenamefont {Ceperley}\ and\ \citenamefont
  {Alder}(1980)}]{ceperley1980ground}%
  \BibitemOpen
  \bibfield  {author} {\bibinfo {author} {\bibfnamefont {D.~M.}\ \bibnamefont
  {Ceperley}}\ and\ \bibinfo {author} {\bibfnamefont {B.}~\bibnamefont
  {Alder}},\ }\href@noop {} {\bibfield  {journal} {\bibinfo  {journal}
  {Physical Review Letters}\ }\textbf {\bibinfo {volume} {45}},\ \bibinfo
  {pages} {566} (\bibinfo {year} {1980})}\BibitemShut {NoStop}%
\bibitem [{\citenamefont {Liu}\ and\ \citenamefont
  {Cohen}(1992)}]{liu1992theoretical}%
  \BibitemOpen
  \bibfield  {author} {\bibinfo {author} {\bibfnamefont {A.~Y.}\ \bibnamefont
  {Liu}}\ and\ \bibinfo {author} {\bibfnamefont {M.~L.}\ \bibnamefont
  {Cohen}},\ }\href@noop {} {\bibfield  {journal} {\bibinfo  {journal}
  {Physical Review B}\ }\textbf {\bibinfo {volume} {45}},\ \bibinfo {pages}
  {4579} (\bibinfo {year} {1992})}\BibitemShut {NoStop}%
\bibitem [{\citenamefont {Sunada}(2008)}]{sunada2008crystals}%
  \BibitemOpen
  \bibfield  {author} {\bibinfo {author} {\bibfnamefont {T.}~\bibnamefont
  {Sunada}},\ }in\ \href@noop {} {\emph {\bibinfo {booktitle} {Notices Amer.
  Math. Soc}}}\ (\bibinfo {organization} {Citeseer},\ \bibinfo {year}
  {2008})\BibitemShut {NoStop}%
\bibitem [{\citenamefont {Perdew}\ \emph {et~al.}(1996)\citenamefont {Perdew},
  \citenamefont {Burke},\ and\ \citenamefont
  {Ernzerhof}}]{perdew1996generalized}%
  \BibitemOpen
  \bibfield  {author} {\bibinfo {author} {\bibfnamefont {J.~P.}\ \bibnamefont
  {Perdew}}, \bibinfo {author} {\bibfnamefont {K.}~\bibnamefont {Burke}}, \
  and\ \bibinfo {author} {\bibfnamefont {M.}~\bibnamefont {Ernzerhof}},\
  }\href@noop {} {\bibfield  {journal} {\bibinfo  {journal} {Physical review
  letters}\ }\textbf {\bibinfo {volume} {77}},\ \bibinfo {pages} {3865}
  (\bibinfo {year} {1996})}\BibitemShut {NoStop}%
\bibitem [{\citenamefont {Bl\"ochl}(1994)}]{PAW_Bloch}%
  \BibitemOpen
  \bibfield  {author} {\bibinfo {author} {\bibfnamefont {P.~E.}\ \bibnamefont
  {Bl\"ochl}},\ }\href {\doibase 10.1103/PhysRevB.50.17953} {\bibfield
  {journal} {\bibinfo  {journal} {Phys. Rev. B}\ }\textbf {\bibinfo {volume}
  {50}},\ \bibinfo {pages} {17953} (\bibinfo {year} {1994})}\BibitemShut
  {NoStop}%
\bibitem [{\citenamefont {Kresse}\ and\ \citenamefont
  {Joubert}(1999)}]{kresse1999ultrasoft}%
  \BibitemOpen
  \bibfield  {author} {\bibinfo {author} {\bibfnamefont {G.}~\bibnamefont
  {Kresse}}\ and\ \bibinfo {author} {\bibfnamefont {D.}~\bibnamefont
  {Joubert}},\ }\href@noop {} {\bibfield  {journal} {\bibinfo  {journal}
  {Physical review b}\ }\textbf {\bibinfo {volume} {59}},\ \bibinfo {pages}
  {1758} (\bibinfo {year} {1999})}\BibitemShut {NoStop}%
\bibitem [{\citenamefont {Limbu}\ \emph {et~al.}(2018)\citenamefont {Limbu},
  \citenamefont {Madueke}, \citenamefont {Atta-Fynn}, \citenamefont {Drabold},\
  and\ \citenamefont {Biswas}}]{limbu2018}%
  \BibitemOpen
  \bibfield  {author} {\bibinfo {author} {\bibfnamefont {D.~K.}\ \bibnamefont
  {Limbu}}, \bibinfo {author} {\bibfnamefont {M.~U.}\ \bibnamefont {Madueke}},
  \bibinfo {author} {\bibfnamefont {R.}~\bibnamefont {Atta-Fynn}}, \bibinfo
  {author} {\bibfnamefont {D.~A.}\ \bibnamefont {Drabold}}, \ and\ \bibinfo
  {author} {\bibfnamefont {P.}~\bibnamefont {Biswas}},\ }\href@noop {}
  {\bibfield  {journal} {\bibinfo  {journal} {arXiv: 1809.00300}\ } (\bibinfo
  {year} {2018})}\BibitemShut {NoStop}%
\end{thebibliography}%

\end{document}